\documentclass[aps,prx,floatfix,superscriptaddress,noshowpacs,twocolumn]{revtex4-2}
\usepackage{graphics,graphicx,epsfig,amsmath,amssymb,epstopdf,wasysym,comment}
\usepackage{hyperref}
\usepackage[dvipsnames]{xcolor}
\usepackage{braket,amsfonts}
\usepackage{dsfont}
\usepackage{hyperref}
\usepackage[dvipsnames]{xcolor}
\usepackage{amsthm}
\theoremstyle{remark}

\newcommand{\be}{\begin{equation}}
\newcommand{\ee}{\end{equation}}
\newcommand{\bga}{\begin{gather}}
\newcommand{\ega}{\end{gather}}
\newcommand{\bea}{\begin{eqnarray}}
\newcommand{\eea}{\end{eqnarray}}
\newcommand{\dagga}{{\phantom{\dagger}}}

\newcommand{\bQ}{\mathbf{Q}}

\newcommand{\bq}{\mathbf{q}}

\newcommand{\bk}{\mathbf{k}}

\newcommand{\Ima}{\text{Im}}
\newcommand{\Rea}{\text{Re}}
\newcommand{\dis}{\displaystyle}

\newcommand{\up}{\uparrow}
\newcommand{\down}{\downarrow}
\newcommand{\fract}[2]{\frac{\dis \;#1\;}{\dis \;#2\;}}
\newcommand{\Tr}{\mathrm{Tr}}
\newcommand{\eqn}[1]{(\ref{#1})}

\newcommand{\ep}{{\epsilon}}

\newcommand{\bw}{\begin{widetext}}
\newcommand{\ew}{\end{widetext}}
\newenvironment{eqs}%
{\begin{equation} \begin{aligned}}%
{\end{aligned} \end{equation} }
\newcommand{\beal}{\begin{eqs}}
\newcommand{\eal}{\end{eqs}}
\newcommand{\bd}[1]{{\boldsymbol{#1}}}
\newcommand{\esp}[1]{\text{e}^{#1}}
\newcommand{\bealn}{\beal\nonumber}

\begin{document}
\title{Luttinger's theorem in presence of Luttinger surfaces}

\author{Jan Skolimowski}
\affiliation{International School for Advanced Studies (SISSA), Via Bonomea 265, I-34136 Trieste, Italy} 
\author{Michele Fabrizio}
\affiliation{International School for Advanced Studies (SISSA), Via Bonomea 265, I-34136 Trieste, Italy} 

\begin{abstract}
\textbf{Breakdown of Landau's hypothesis of adiabatic continuation from non-interacting to fully interacting electrons is commonly believed to bring about a violation of Luttinger's theorem. Here, we elucidate what may go wrong in the 
proof of Luttinger's theorem. The analysis provides a simple 
way to correct Luttinger's expression of the electron number in single-band models where perturbation theory breaks down through the birth of a Luttinger surface without symmetry breaking. In those cases, we find that the Fermi volume only accounts for the doping 
away from half-filling. In the hypothetical circumstance of a non-symmetry breaking Mott insulator 
with a Luttinger surface, our analysis predicts the noteworthy existence of quasiparticles whose `Fermi` surface 
is just the Luttinger one. Therefore, those quasiparticles can be legitimately regarded as `spinons`, and the Mott insulator 
with a Luttinger surface as  realisation of a spin-liquid insulator.}

\end{abstract}

                         
\maketitle

\section{Introduction}

Landau originally derived his celebrated theory of Fermi liquids~\cite{Landau1,*Landau2} assuming that the non-interacting 
many-body excited states evolve adiabatically into the fully interacting ones 
upon gradually turning on interaction. 
The theory was later microscopically derived~\cite{Nozieres&Luttinger-1,*Nozieres&Luttinger-2} by means of the just developed diagrammatic many-body 
perturbation theory~\cite{Luttinger&Ward,Luttinger-1}. A famous by-product of the diagrammatic formalism is the so-called Luttinger theorem~\cite{Luttinger}, which, in conventional Landau's 
Fermi liquids, predicts that the volume fraction enclosed by the quasiparticles' Fermi surface is just the electron filling fraction.  Over the years, Landau's adiabatic hypothesis and Luttinger's theorem have become 
almost synonyms, in the sense that if one is violated, so is the other.  Such belief has been mostly triggered  by the anomalous properties of many strongly correlated materials, especially underdoped 
cooper-oxide superconductors.\\
However,  the traditional derivation~\cite{Luttinger,AG&D,Langer-PR1961,Langreth-PRB1975} of Luttinger's theorem simply relies on the existence of a Luttinger-Ward functional~\cite{Luttinger&Ward}, which can be constructed non-perturbatively~\cite{Potthoff-2006}.   
Therefore, it is not at all evident why Luttinger's theorem should be violated at the breakdown of perturbation theory,
as it is likewise not true that Landau's Fermi liquid theory applies only in the perturbative regime~\cite{mio,mio-2}.\\ 
The topological arguments by Oshikawa~\cite{Oshikawa-PRL2000} in periodic models clarify Luttinger's theorem violation in non-trivial examples that host 
fractionalised excitations~\cite{Fractionalised-Senthil-PRL2003,Senthil-PRB2004} 
or topological order~\cite{Vishwanath-PRB2004,Hastings_2005}, but does not allow 
identifying at which point the traditional proof may go wrong. Moreover, it is not instructive in non-magnetic Mott insulators at integer number of electrons per site, 
like the model discussed by Rosh~\cite{Rosh-2007}, where Luttinger's theorem 
does not yield the correct electron number, nor in models that lack translation symmetry, like quantum impurity models, where Luttinger's theorem is still applicable and can be violated~\cite{Hewson-2018}. \\
The detailed analysis of Heath and Bedell~\cite{Heath_2020} highlights which properties the self-energy must possess for Luttinger's theorem to hold true, even 
in non-periodic models. However, it leaves open the question how to count the number of particles when Luttinger's theorem is violated. 
\\ 
Indeed, there are by now several examples of Luttinger's theorem violation, 
see, e.g., Refs.~\cite{Altshuler-EPL1998,Georges-PRB2001,Fractionalised-Senthil-PRL2003,Senthil-PRB2004,Vishwanath-PRB2004,Rosh-2007,Phillips-PRL2013,Hewson-2018}.
In addition, there are numerical evidences that 
Luttinger's theorem fails in models of doped Mott insulators below a critical doping~\cite{Kotliar-PRB2006,Prelovsek-PRB2007,Becca-PRB2012,Georges-PNAS2018,Georges-PRX2018,Trivedi-PRB2021} that seems to be associated with the birth of a Luttinger surface~\cite{Igor-PRB2003}, which, according to Ref.~\cite{Heath_2020}, does violates the requirements for Luttinger's theorem validity. \\ 
In this work, we try to shed further light on such fundamental issue, beyond the 
great progresses that have been already accomplished~\cite{Fractionalised-Senthil-PRL2003,Vishwanath-PRB2004,Heath_2020,Else-PRX2021,Wen-PRB2021}. We do that paying particular attention to the role of Luttinger surfaces, or, more generally, to the zeros of the single-particle Green's function at zero imaginary frequency, a concept that does not require translation invariance.   

\section{Luttinger's theorem}
\label{Luttinger's theorem}
We start by deriving Luttinger's theorem in a slightly different way as conventionally done~\cite{Luttinger,AG&D}, 
somehow closer to Refs.~\cite{Langer-PR1961,Langreth-PRB1975}, which better highlights under which circumstances that theorem may fail. 
Moreover, the derivation below, though based on old-fashioned many-body theory, 
naturally brings to the concept of 'quasiparticles', and their Fermi or Luttinger surfaces~\cite{mio-2}. \\

\noindent
We consider a system of interacting electrons with annihilation operators 
$c^\dagga_\alpha$ corresponding to a complete basis of single-particle wavefunctions 
labelled by $\alpha=1,\dots,K$, with $K\to\infty$ in the thermodynamic limit. The Hamiltonian admits a set of conserved quantities 
$Q$, represented by hermitian matrices $\hat{Q}$ with components $Q_{\alpha\beta}$ defined 
in such a way that the eigenvalues are integers.  
$Q_{\alpha\beta}=\delta_{\alpha\beta}$ corresponds to the total number $N$ of electrons, while all other independent $Q$'s are represented by traceless 
matrices $\hat Q$. We hereafter consider the evolution of the operators in imaginary time and use the Matsubara formalism, which is more convenient~\cite{Igor-PRB2003} since on the imaginary frequency axis the single particle Green's function and self-energy cannot have singularities but, eventually, at the origin. 
Moreover, to avoid any issue related to the discontinuity at zero imaginary time of the Green's functions, we use instead of $N$ the deviation $N-K/2$ of the 
electron number with respect to half-filling, so that we can write the expectation value of any conserved quantity as  
\beal
Q &= \fract{1}{2}\,\sum_{\alpha\beta}\, Q_{\beta\alpha}\,\Big(
\langle\,c^\dagger_\beta\,c^\dagga_\alpha\,\rangle - \langle\,c^\dagga_\alpha\,c^\dagger_\beta\,\rangle\Big)\\
&= T\,\sum_{n}\,\Tr\Big(\hat{G}(i\ep_n)\,\hat{Q}\Big)\,,\label{Q1}
\eal
where $\hat{G}(i\ep_n)=\hat{G}(-i\ep_n)^\dagger$ is the Green's function matrix in  Matsubara frequencies 
$\ep_n=(2n+1)\,\pi T$. According to Dyson's equation, 
\beal
\hat{G}^{-1}(i\ep_n) &= i\ep_n\,\hat{I} -\hat{H}_0 -\hat{\Sigma}(i\ep_n)\,,\label{Dyson}
\eal 
with $\hat{I}$ the identity matrix, and $\hat{H}_0$ the non-interacting Hamiltonian, including the chemical potential term, represented in the chosen 
basis. 
$\hat{\Sigma}(i\ep_n)=\hat{\Sigma}(-i\ep_n)^\dagger$ is the 
self-energy matrix that accounts for all interaction effects. 
We can equivalently write Eq.~\eqn{Q1} as
\beal
Q &= -T\sum_n \fract{\partial}{\partial i\ep_n}\,
\Tr\Big(\ln\hat{G}(i\ep_n)\;\hat{Q}\Big) + I_L(Q)\,,\label{Q}
\eal
where
\beal
I_L(Q) &=  T\sum_n\,\Tr\bigg(\hat{G}(i\ep_n)\,\fract{\partial\hat{\Sigma}(i\ep_n)}
{\partial i\ep_n}\;\hat{Q}\bigg)
\,.\label{I_L}
\eal
Hereafter, we denote Eq.~\eqn{I_L} as the Luttinger integral for the conserved quantity $Q$,  and use simply $I_L$ for 
the case $\hat{Q}=\hat{I}$.  \\
We just note that at particle-hole symmetry $I_L(Q)$ vanishes identically for all non particle-hole invariant $Q$'s, thus 
also the total electron number, in which case Luttinger's theorem holds trivially.
Seemingly, $I_L(Q)=0$ in absence of interaction, where $\hat{\Sigma}(i\ep_n)=0$.\\ 
In more general circumstances, we consider the Luttinger-Ward functional $\Phi[G]$ 
satisfying~\cite{Luttinger&Ward,Potthoff-2006}
\beal
\delta\Phi[G] &= T\,\sum_n\,\esp{i\ep_n\eta}\,
\Tr\Big(\hat{\Sigma}(i\ep_n)\,\delta\hat{G}(i\ep_n)\Big)\,,\label{Phi[G]-1}
\eal
with $\eta>0$ that must be sent to zero after performing the summation.
In perturbation theory, the explicit expression of $\Phi[G]$ reads~\cite{Luttinger&Ward}
\beal
\Phi[G] &= T\sum_n\,\esp{i\ep_n\eta}\, 
\sum_{m\geq 1}\fract{1}{2m}\,\Tr\Big(\hat{G}(i\ep_n)\,\hat{\Sigma}^{(m)}(i\ep_n)\Big)\\
&\equiv T\sum_n\,\esp{i\ep_n\eta}\,\Phi(i\ep_n)\,,\label{Phi[G]-2}
\eal
where $\hat{\Sigma}^{(m)}(i\ep_n)$ is the sum of all $m$-th order skeleton diagrams.
We assume that the non-perturbative $\Phi[G]$~\cite{Potthoff-2006} can be still written as a series of 
terms $\Phi(i\ep_n)$ as in Eq.~\eqn{Phi[G]-2}. Through Eqs.~\eqn{Phi[G]-1} and \eqn{Phi[G]-2} it readily follows that
\beal
\fract{\delta\Phi[G]}{\delta i\ep}&\equiv T\,\sum_n\,
\Tr\left(\hat{\Sigma}(i\ep_n)\,\fract{\partial\hat{G}(i\ep_n)}{\partial i\ep_n}\right)\\
&= T\sum_n\,\fract{\partial\Phi(i\ep_n)}{\partial i\ep_n}\;,\label{Phi[G]-3}
\eal
where we set $\eta=0$ before performing the sum since the 
function decays faster than $1/\ep_n$ for $\ep_n\to\pm\infty$.
Eq.~\eqn{Phi[G]-3} allows us to rewrite $I_L(Q)$ of Eq.~\eqn{I_L} for $\hat{Q}=\hat{I}$ simply 
as
\beal
I_L &= T\sum_n\,\Tr\bigg(\hat{G}(i\ep_n)\,\fract{\partial\hat{\Sigma}(i\ep_n)}
{\partial i\ep_n}\bigg)\\
&=T\sum_n\,\fract{\partial I_L(i\ep_n)}{\partial i\ep_n}\;,\label{I_L-deriv}
\eal
where
\beal
I_L(i\ep_n) &= \Tr\Big(\hat{\Sigma}(i\ep_n)\,\hat{G}(i\ep_n)\Big)
-\Phi(i\ep_n)\,.
\eal
In other words, it is always possible to represent the Luttinger integral as a sum over $\ep_n$ of a derivative. It follows that the total number of electrons can be written as 
\bealn
N &=  \fract{K}{2} -T\sum_n \;\fract{\partial}{\partial i\ep_n}\,
\Tr\Big(\ln\hat{G}(i\ep_n)\Big) \\
&\qquad + T\sum_n\,\fract{\partial I_L(i\ep_n)}{\partial i\ep_n}\\
&\;\xrightarrow[T\to 0]{} 
\fract{K}{2} -\int_{-\infty}^\infty \fract{d\ep}{2\pi}\;\fract{\partial}{\partial i\ep}\,
\Tr\Big(\ln\hat{G}(i\ep)\Big) \\
&\qquad + \int_{-\infty}^\infty \fract{d\ep}{2\pi}\;\fract{\partial I_L(i\ep)}{\partial i\ep}\;.
\eal
Since $\hat{G}(-i\ep)=\hat{G}(i\ep)^\dagger$ and, similarly, $I_L(-i\ep)=I_L(i\ep)^*$, if we define, through 
the polar decomposition of  $\hat{G}(i\ep)$, the matrix  
\beal
\hat{\delta}(\ep) &\equiv \arg\big(\hat{G}(i\ep)\big) = \Ima\,\ln \big(\hat{G}(i\ep)\big)\,,\label{LT: hat-delta}
\eal
then, for $T\to 0$,  and noticing that $\Ima\,I_L(i\ep)\to 0$ while $\hat{\delta}(\ep)\to -\pi/2\; \hat{I}$
for $\ep\to\infty$, 
\beal
N &= \fract{K}{2} + \int_{-\infty}^\infty \fract{d\ep}{2\pi}\;\Tr\Big(\hat{G}(i\ep)\Big)\\
&=
K  + \fract{1}{\pi}\;\Tr\Big(\hat{\delta}(0^+)\Big) 
-\fract{1}{\pi}\; \Ima\,I_L(i0^+)\,.
\label{LT: general statement}
\eal
This expression is exact. It is still not Luttinger's theorem but a kind of generalisation of it, and it is remarkable 
as it shows that a quantity requiring integration over all frequencies can be alternatively calculated through boundary terms. \\
In reality, Luttinger's theorem statement 
is that $\Ima\,I_L(i0^+)=0$ in Eq.~\eqn{LT: general statement}, which is not to be expected a priori. Nonetheless, the proof goes as follows. The Luttinger-Ward functional $\Phi[G]$ is invariant if the Matsubara frequency of each internal Green's function is replaced, see Eq.~\eqn{Dyson}, by $i\ep_n\,\hat{I}+ i\omega\,\hat{Q}$ for any 
conserved $Q$, where $\omega=2\pi\,T$. Therefore, 
\be
0= \fract{\Delta^Q\Phi[G]}{i\omega}= T\sum_n\,
\Tr\bigg(\hat{\Sigma}(i\ep_n)\,
\fract{\Delta^Q \hat{G}(i\ep)}{i\omega}\bigg),\label{delta Phi = 0}
\ee
with
\beal
\fract{\Delta^Q\hat{G}(i\ep)}{i\omega}
\equiv \fract{\hat{G}(i\ep_n+i\omega\,\hat{Q})-\hat{G}(i\ep_n)}{i\omega}\;,
\eal
the finite difference of $\hat{G}(i\ep)$. For $\hat{Q}=\hat{I}$ that implies 
\beal
0&= T\sum_n\,
\Tr\bigg(\hat{\Sigma}(i\ep_n)\,
\fract{\;\hat{G}(i\ep_n+i\omega)-\hat{G}(i\ep_n)\;}{i\omega}\bigg)\\
&= - T\sum_n\,
\Tr\bigg(\hat{G}(i\ep_n)\,
\fract{\;\hat{\Sigma}(i\ep_n+i\omega)-\hat{\Sigma}(i\ep_n)\;}{i\omega}\bigg)\\
&\equiv -T\sum_n\;\fract{\;I_L(i\ep_n+i\omega)-I_L(i\ep_n)\;}{i\omega}\equiv -I^\Delta_L\;,
\label{finite-difference}
\eal
which just means that the convergence of the series allows the change of variable $i\ep_n+i\omega\to i\ep_n$ 
that makes $I^\Delta_L$ trivially vanish. It is tempting to assume that $I^\Delta_L$, i.e., 
the sum over $\ep_n$ of the finite difference, coincides with $I_L$ in Eq.~\eqn{I_L-deriv}, i.e., the sum 
over $\ep_n$ of the derivative, in the limit $T\to 0$, thus $\omega\to 0$. That is actually what is commonly assumed in the proof of Luttinger's theorem, 
in which case $I_L=0$ follows, and thus $\Ima\,I_L(i0^+)=0$ in Eq.~\eqn{LT: general statement}. 
However, that apparently reasonable assumption is not at all guaranteed, as we now discuss. \\
In the Supplementary Notes of Ref.~\cite{mio-2} it has been shown that, at leading order 
in $T$, 
\beal
I_L & =-\fract{1}{\pi}\,\Ima\,I_L(i0^+)= I_L -I^\Delta_L\\
&\simeq -\fract{1}{4\pi i}\,\lim_{\ep\to 0^+}\,S(i\ep)
\,.\label{LT: condition}
\eal
where 
\beal
S(i\ep) \equiv \Tr\bigg[ \Big(\hat{G}(i\ep)+ \hat{G}(i\ep)^\dagger\Big)
\,\Big( \hat{\Sigma}(i\ep) - \hat{\Sigma}(i\ep)^\dagger\Big)\bigg]\,.
\label{LT: def S(iep)}
\eal
It follows that, if $S(i\ep)$ is finite for $\ep\to 0^+$, then Luttinger's theorem is definitely violated. That happens, e.g., 
in the Sachdev-Ye-Kitaev model~\cite{Sachdev&Ye-PRL1993,Georges-preprint2021}. 
On the contrary, one can 
readily prove that $S(i\ep\to 0^+)=0$ when perturbation theory holds. Indeed, if we define the `quasiparticle` residue
\beal
\sqrt{\,\hat{Z}(i\ep)^\dagger{^{-1}}\,\hat{Z}(i\ep)^{-1}\;} &\equiv 
\hat{I} -\fract{\;\hat{\Sigma}(i\ep) - \hat{\Sigma}(i\ep)^\dagger\;}{2i\ep}\;,
\label{LT: def Z}
\eal
where $\hat{Z}(i\ep)=\hat{Z}(-i\ep)$, we do know that perturbatively 
$\hat{Z}(0)=\hat{Z}(0)^\dagger$ is positive definite, so that 
\bealn
\hat{\Sigma}(i\ep) - \hat{\Sigma}(i\ep)^\dagger
\xrightarrow[\ep\to 0]{} 2\,\Big(\hat{I} -\hat{Z}(0)^{-1}\Big)\,i\ep\,,
\eal
and thus $S(i\ep)$ vanishes as $\ep\to 0^+$. However, $S(i\ep\to 0^+)=0$, though necessary for $I_L=0$, is not 
a sufficient condition. The reason is that the right hand side of Eq.~\eqn{LT: condition} is just the leading term of an expansion in $T$. Its vanishing means 
that each term of the series expansion goes to zero as $T\to 0$, which does not guarantee that the whole series vanishes~\cite{mio-2}. In other words, while we can safely 
state that, in the regime where perturbation theory is valid, $S(i\ep\to 0^+)=0$ does imply that $I_L=0$, and thus that Luttinger's theorem holds true, we cannot exclude that the theorem is violated when perturbation theory
breaks down.\\
However, let us assume the necessary condition $S(i\ep\to i0^+)= 0$ and draw its consequences. 
By definition, the single-particle density of states $A$ at the chemical potential is
\bealn
A &= -\lim_{\ep\to 0^+}\,\fract{1}{2\pi i}\,\Tr\Big(\hat{G}(i\ep) - \hat{G}(i\ep)^\dagger\Big)\\
&\equiv \lim_{\ep\to 0^+}\,\Tr\Big(\hat{A}(i\ep)\Big)\,,
\eal
where $\hat{A}(i\ep) = \hat{A}(i\ep)^\dagger=-\hat{A}(-i\ep)$. Through $\hat{A}(i\ep)$, we can write
\bealn
\hat{\Sigma}(i\ep) - \hat{\Sigma}(i\ep)^\dagger &= 2i\ep -2\pi i\,\hat{G}(i\ep)^{-1}\,\hat{A}(i\ep)\,\hat{G}(i\ep)^\dagger{^{-1}}\,,
\eal
and thus $S(i\ep)$ in Eq.~\eqn{LT: def S(iep)} becomes 
\bealn
S(i\ep) &= 2i\ep\, \Tr\Big(\hat{G}(i\ep)+ \hat{G}(-i\ep)\Big)\\
&\quad -2\pi i\, \Tr\bigg[ \Big(\hat{G}(i\ep)^{-1}+ \hat{G}(i\ep)^\dagger{^{-1}}\Big)
\,\hat{A}(i\ep)\bigg]\,.
\eal
We now formally filter out the 'quasiparticle' Green's function through the 'quasiparticle' residue Eq.~\eqn{LT: def Z}, 
\beal
\hat{G}_\text{qp}(i\ep)^{-1} &\equiv \sqrt{\,\hat{Z}(i\ep)^\dagger\;}\;
\hat{G}(i\ep)^{-1}\;
\sqrt{\,\hat{Z}(i\ep)\;} \\
&= i\ep\,\hat{I}-\hat{\Xi}(i\ep) \,,
\label{LT: def quasiparticle Green}
\eal
where  
\beal
\hat{\Xi}(i\ep) &\equiv  \sqrt{\,\hat{Z}(i\ep)^\dagger\;}\;\Big(\hat{H}_0 +\Rea\,\hat{\Sigma}(i\ep)\Big)\;
\sqrt{\,\hat{Z}(i\ep)\;}\,,
\label{LT: def Xi}
\eal
is a $K\times K$ hermitian matrix, and thus has real eigenvalues 
$\ep_{*\ell}(\ep)=\ep_{*\ell}(-\ep)$, $\ell=1,\dots,K$. 
Therefore, if we further define 
\beal
&\hat{A}_\text{qp}(i\ep)  \equiv -\fract{1}{2\pi i}\,\Big(\hat{G}_\text{qp}(i\ep) - \hat{G}_\text{qp}(i\ep)^\dagger\Big) \\
&\quad = \fract{\ep}{\pi }\,
\hat{G}_\text{qp}(i\ep)\,\hat{G}_\text{qp}(i\ep)^\dagger
= \fract{\ep}{\pi }\,\fract{1}{\;\ep^2 + \hat{\Xi}(i\ep)^2\;} \\
&\quad= 
\sqrt{\,\hat{Z}(i\ep)^{-1}\;}\;\hat{A}(i\ep)\;\sqrt{\,\hat{Z}(i\ep)^\dagger{^{-1}}\;}\,,
\label{LT: quasiparticle A}
\eal
which is diagonal in the basis that diagonalises $\hat{\Xi}(i\ep)$ with elements 
\bealn
A_{\text{qp}\,\ell}(i\ep) = \fract{1}{\pi}\;\fract{\ep}{\;\ep^2+ \ep_{*\ell}(\ep)^2\;}\;,
\eal
then
\beal
S(i\ep) &= 2i\ep\, \Tr\Big(\hat{G}(i\ep)+ \hat{G}(-i\ep)\Big)\\
&\qquad +4\pi i\, \Tr\bigg[\, \hat{\Xi}(i\ep)\,\hat{A}_\text{qp}(i\ep)\,\bigg]\\
&= 2i\ep\, \Tr\Big(\hat{G}(i\ep)+ \hat{G}(-i\ep)\Big)\\
&\quad +4\pi i\,\sum_{\ell=1}^K\, \fract{\;\ep_{*\ell}(\ep)\;}{\pi}\;\fract{\ep}{\;\ep^2+\ep_{*\ell}(\ep)^2\;}\;.
\label{LT: def S}
\eal
Since the first term on the right hand side of Eq.~\eqn{LT: def S} vanishes for $\ep\to0$, the necessary condition for Luttinger's theorem to hold becomes
\beal
&\lim_{\ep\to 0^+}\,  \Tr\Big[\, \hat{\Xi}(i\ep)\,\hat{A}_\text{qp}(i\ep)\,\Big]\\
&\qquad = \lim_{\ep\to 0^+}\,\sum_{\ell=1}^K\, \fract{\;\ep_{*\ell}(\ep)\;}{\pi}\;\fract{\ep}{\;\ep^2+\ep_{*\ell}(\ep)^2\;} 
= 0\,.
\label{LT: condition 2}
\eal
In the thermodynamic limit, $K\to\infty$, $\ep_{*\ell}(\ep)$ defines a continuous spectrum 
where $\ell$ runs in a $d$-dimensional space, with $d$ the spatial dimension of the system times the number of internal degrees of freedom. For instance, 
in the periodic case, $\ell$ labels the momentum within the Brillouin zone, the band index and the spin. 
Any $\ell$ such that $\ep_{*\ell}(\ep\to 0)\not= 0$ yields a contribution to the sum \eqn{LT: condition 2} 
that trivially vanishes as $\ep\to 0$. Let us instead consider the manifold $\ell=\ell_*$ such that 
$\ep_{*\ell_*}(\ep\to 0)= 0$. If, for a given $\ell_*$, $\ep_{*\ell_*}(\ep\to 0)\sim c_*\,|\ep|^\alpha$, with $\alpha>0$, 
its contribution to the sum \eqn{LT: condition 2} is
\bealn
&\fract{\;\ep_{*\ell_*}(\ep)\;}{\pi}\;\fract{\ep}{\;\ep^2+\ep_{*\ell_*}(\ep)^2\;} \\
&\qquad \xrightarrow[\ep\to 0^+]{} \fract{\;c_*\,|\ep|^\alpha\;}{\pi}\;\fract{\ep}{\;\ep^2+c_*^2\,|\ep|^{2\alpha}\;} \;,
\eal
and vanishes only if $\alpha >1$, which thus becomes the necessary condition for the validity of Luttinger's theorem. 
We can further distinguish two different cases. For instance, if we assume that 
\beal
\mathbf{\bullet}\; \hat{\Xi}(i\ep)~\text{is,~at~leading~order,~analytic~at~}\ep=0\,,
\label{LT: condition quasiparticles}
\eal 
then $\alpha =2$ since $\ep_{*\ell}(\ep)$ is even in $\ep$, which automatically satisfies the necessary condition for Luttinger's theorem to hold. 
In this case, $\ep_{*\ell}(\ep\to 0) \simeq \ep_{*\ell}(0)+ O\big(\ep^{2}\big)$,  where 
$\ep_{*\ell}(0)\equiv \ep_{*\ell}$ are the eigenvalues of 
\beal
\hat{H}_* &\equiv  \sqrt{\,\hat{Z}(0)^\dagger\;}\,\Big(\hat{H}_0 +\hat{\Sigma}(0)\Big)\,
\sqrt{\,\hat{Z}(0)\;}\,.
\label{LT: def H_*}
\eal
Accordingly, the `quasiparticle` Green's function and density of states at the chemical potential are 
\beal
\hat{G}_\text{qp}(i\ep) &\xrightarrow[\ep\to 0]{}
\fract{1}{\;i\ep\,\hat{I} -\hat{H}_* \;}\;,\\
A_\text{qp} &= \lim_{\ep\to 0^+}\,\Tr\Big(\hat{A}_\text{qp}(i\ep)\Big)
= \sum_\ell\,\delta\big(\ep_{*\ell}\big)\,,
\label{LT: quasiparticle conjecture}
\eal
and correspond to those 
of free particles, thus the `quasiparticles', described by the `quasiparticle` Hamiltonian 
$\hat{H}_*$ with eigenvalues $\ep_{*\ell}$. \\
On the contrary, if $\hat{\Xi}(i\ep)$ is non analytic and yet satisfies the necessary condition for Luttinger's theorem, then $1<\alpha<2$, since any non-analyticity yielding non-integer $\alpha>2$ will be hidden by 
the ever-present analytical terms. That is precisely what happens for interacting electrons in one dimension. 
Those systems do not sustain quasiparticles 
in the sense of Eq.~\eqn{LT: quasiparticle conjecture}, and yet Luttinger's theorem 
is valid~\cite{Bedell-PRL1997,Affleck-PRL1997}. The same occurs in marginal Fermi liquids~\cite{Varma-PRL1989}, or metals with quantum critical behaviour~\cite{Chubukov-AdvPhys2003}, which, despite a 
non-analytic self-energy, satisfy Luttinger's theorem~\cite{Heath_2020}. 
Conversely, since $S(i\ep\to i0^+)=0$ is not sufficient for Luttinger's theorem to hold, we must also conclude that `quasiparticles` may exist even when Luttinger's theorem is violated~\cite{mio-2}.\\
We also emphasise that $1<\alpha<2$ entails singularities in perturbation theory. Therefore, Eq.~\eqn{LT: condition quasiparticles} must be always verified when  perturbation theory is well defined, which is equivalent to saying that quasiparticles always exist in the perturbative regime, in agreement with Landau's adiabatic hypothesis.  
\\

Hereafter, we assume the analyticity condition \eqn{LT: condition quasiparticles}, thus Eq.~\eqn{LT: quasiparticle conjecture}. We believe that this choice, though limiting, may be pertinent to doped Mott insulators in dimensions $d>1$~\cite{Kotliar-PRB2006,Prelovsek-PRB2007,Becca-PRB2012,Georges-PNAS2018,Georges-PRX2018,Trivedi-PRB2021}. In that case, $\hat{\delta}(\ep)$ 
is diagonal in the basis 
that diagonalises $\hat{H}_*$ with elements $-\pi + \pi\,\theta\big(-\ep_{*\ell}\big)$. It follows that Eq.~\eqn{LT: general statement} 
becomes
\bealn
N &= 
K  + \fract{1}{\pi}\;\Tr\Big(\hat{\delta}(0^+)\Big) 
-\fract{1}{\pi}\; \Ima\,I_L(i0^+)\\
&= \sum_{\ell=1}^K\,\theta\big(-\ep_{*\ell}\big)-\fract{1}{\pi}\; \Ima\,I_L(i0^+)\,,
\eal
which represents the general statement \eqn{LT: general statement} of Luttinger's theorem when `quasiparticles` exist. We note that $N$ is integer at $T=0$ and so is the sum over $\ell$, which implies that the Luttinger integral $I_L$ is 
quantised in integer values when $\eqn{LT: condition quasiparticles}$ holds. 
Therefore, 
\beal
N &= \sum_{\ell=1}^K\,\theta\big(-\ep_{*\ell}\big) + \mathcal{L}\,,& \mathcal{L} &\in\mathbb{Z}\,,
\label{LT: conventional and unconventional statement}
\eal
where $\mathcal{L}=0$ in the perturbative regime, in which case conventional Luttinger's theorem holds, while $\mathcal{L}$ may be finite when perturbation theory breaks down.

\subsection{Generalised Luttinger's theorem in presence of quasiparticles and in periodic systems}
\label{Generalised Luttinger's theorem in presence of quasiparticles and in periodic systems}

In a single-band periodic system invariant under  spin $SU(2)$ symmetry, we have the possibility to further elaborate on the meaning of `quasiparticle`. In this case, $\hat{G}(i\ep)$ 
is diagonal in momentum and spin with elements $G(i\ep,\bk)$ independent of spin, 
and thus $\hat{\Xi}(i\ep)$ is diagonal, too, with elements $\ep_*(\ep, \bk)$ equal for spin $\sigma=\up$ 
and $\down$, now defined, see Eq.~\eqn{LT: def Xi}, as
\be
\ep_*(\ep, \bk) = \big|Z(i\ep,\bk)\big|\;\Big(\ep(\bk)+\Rea\,\Sigma(i\ep,\bk)\Big)
\,.
\label{LT: def ep_*(ep,k)}
\ee
Correspondingly, the quasiparticle, $A_\text{qp}$, and physical electron, $A$, 
density of states at the chemical potential are, in units of the number of sites $V$, 
see Eq.~\eqn{LT: quasiparticle conjecture},
\beal
A_\text{qp} &= \fract{1}{V}\,\sum_{\bk\sigma}\,\delta\big(\ep_*(\bk)\big)\,,\\
A &= \fract{1}{V}\,\sum_{\bk\sigma}\,Z(i\ep\to i0^+,\bk)\,\delta\big(\ep_*(\bk)\big)\,,
\label{LT: quasiparticle and physical particle DOS}
\eal
where $\ep_*(\bk)=\ep_*(\ep\to 0, \bk)$. 
We already know that Eqs.~\eqn{LT: condition 2} and \eqn{LT: condition quasiparticles} 
imply that, if a manifold $\bk=\bk_*$ exists 
such that $\ep_*(0, \bk_*)=0$, then $\ep_*(\ep\to 0, \bk_*)\simeq  \ep^2$. 
We observe that $\ep_*(0, \bk_*)=0$ may occur
\begin{description}
\item[Fermi Surface] if $\bk_*=\bk_F$, with $\bk_F$ such that $\ep(\bk_{F})+\Sigma(0,\bk_F)=0$ while $0<Z(0,\bk_F)<1$, which defines 
a conventional Fermi surface $\bk=\bk_F$ through the roots of $G(0,\bk)^{-1}$ in momentum space. The Fermi surface contribution to the physical electron DOS 
Eq.~\eqn{LT: quasiparticle and physical particle DOS} is finite since $Z(0,\bk_F)\not=0$.

\item[Luttinger Surface] if $\bk_*=\bk_L$, with $\bk_L$ such that $\ep(\bk_{L})+\Sigma(0,\bk_L)\not=0$ but 
\beal
\lim_{\ep\to 0^+}\,\big|Z(i\ep,\bk_L)\big| &=
\lim_{\ep\to 0^+}\,\fract{\ep}{\;\ep -\Ima\,\Sigma(i\ep,\bk_L)\;}
\\ 
&\sim \lim_{\ep\to 0}\,\ep^2 = 0\,,
\label{LT: Luttinger surface}
\eal
which implies $\Sigma(i\ep,\bk_L)\sim 1/i\ep$ and, correspondingly,  $G(i\ep,\bk_L)\to 0$ as $\ep\to 0$. 
Therefore, Eq.~\eqn{LT: Luttinger surface} defines the so-called Luttinger surface~\cite{Igor-PRB2003}, i.e., 
the manifold of roots $\bk=\bk_L$ of $G(0,\bk)$ in momentum space,
whose existence is due to a singular self-energy and thus signals the breakdown of 
perturbation theory. Remarkably, even though the Luttinger surface contribution to the quasiparticle DOS, $A_\text{qp}$ in 
Eq.~\eqn{LT: quasiparticle and physical particle DOS}, is finite, its contribution 
to the physical electron DOS vanishes~\cite{mio-2}.
\end{description}
Therefore, under the analyticity assumption \eqn{LT: condition quasiparticles}, Fermi and Luttinger surfaces 
are both defined by the one and only equation $\ep_*(0, \bk_{F/L})=0$~\cite{mio-2}. 
Moreover, as we earlier mentioned, if perturbation theory is valid there are always quasiparticles, only a Fermi surface may exist within the Brillouin zone, 
and, see Eq.~\eqn{LT: conventional and unconventional statement} at $\mathcal{L}=0$,   
\beal
N &= \sum_{\bk\sigma}\,\theta\big(-\ep_*(\bk)\big)\,,
\label{LT: conventional statement periodic}
\eal
which is the standard perturbative Luttinger's theorem statement that the fraction of the quasiparticle Fermi volume, i.e., the manifold of $\bk:\,\ep_*(\bk)<0$, with respect to the whole Brillouin zone is equal to the electron filling fraction 
$\nu=N/2V$.\\
When perturbation theory breaks down without breaking translational and spin $SU(2)$ 
symmetries, and Luttinger surfaces appear inside the Brillouin zone, we must use the more general formula 
\beal
N &= \sum_{\bk\sigma}\,\theta\big(-\ep_*(\bk)\big) + \mathcal{L}\,,& \mathcal{L}&\in\mathbb{Z}\,,
\label{LT: unconventional statement periodic}
\eal
and thus the quasiparticle Fermi volume fraction no more accounts for the electron filling fraction.\\

\begin{figure}[thb]
\centerline{\includegraphics[width=0.4\textwidth]{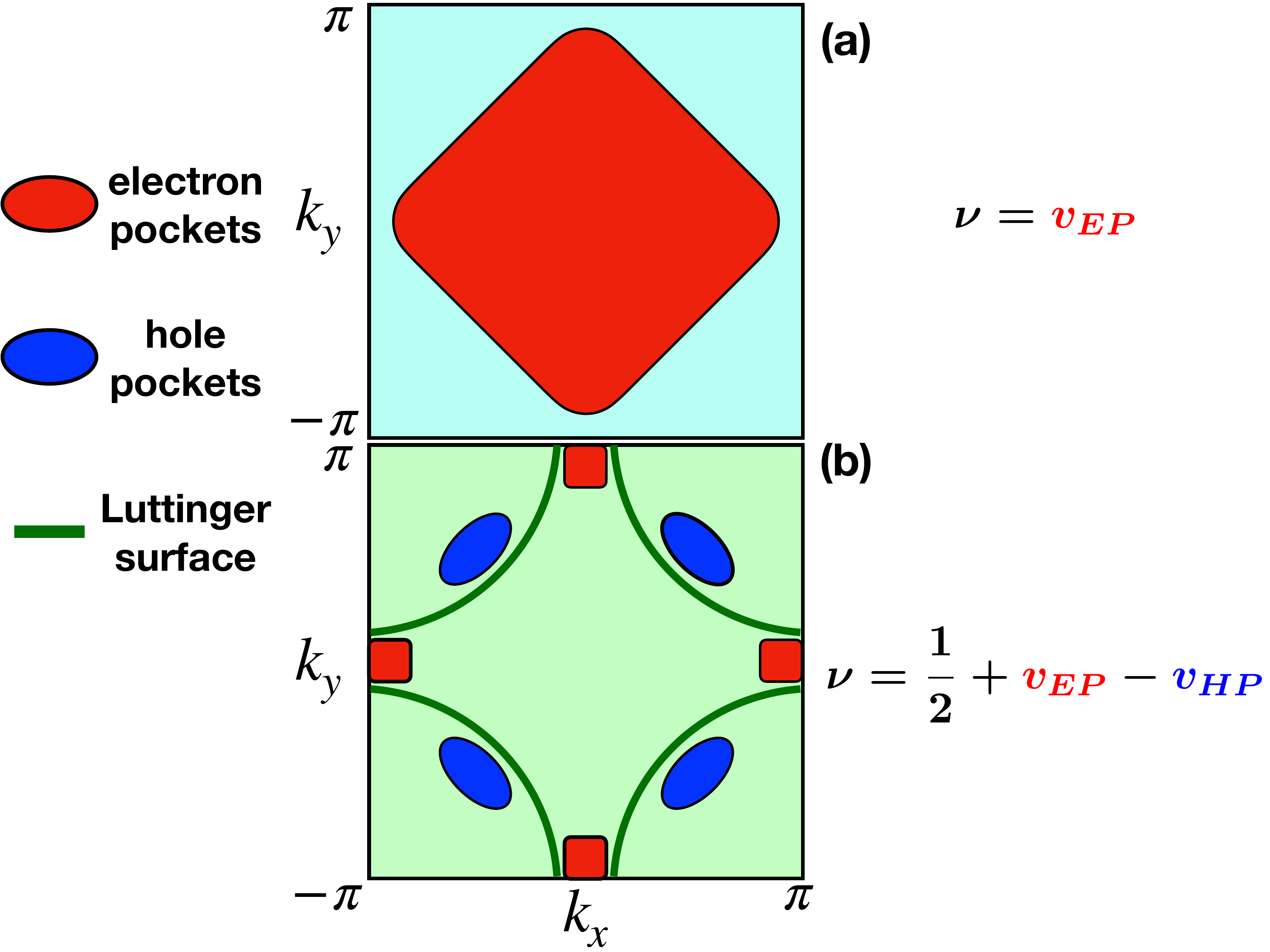}}
\caption{Graphical representation of electron count when perturbation theory is valid, panel (a), or, panel (b), when it breaks down and a Luttinger surface appears,  green line in the figure. Electron-like, i.e., $\ep_*(\bk)<0$, and hole-like, i.e., $\ep_*(\bk)>0$, Fermi pockets are shown, respectively, in red and blue and have volume fraction $v_{EP}$ and $v_{HP}$ with respect to the whole Brillouin zone. When perturbation theory is valid, the electron filling fraction $\nu=N/2V$, where 
$N$ is the total number of electrons and $V$ the number of sites, is simply given by $\nu=v_{EP}$, panel (a). When a Luttinger surface exists, the filling fraction is obtained through $\nu=1/2 +v_{EP}-v_{HP}$, panel (b). }
\label{count}
\end{figure}

\noindent
In order to proceed in this case, we use Oshikawa's topological approach to Luttinger's theorem in periodic systems~\cite{Oshikawa-PRL2000}.  We first note that the above 'quasiparticle' derivation holds even when 
the system is a non-symmetry breaking Mott insulator provided it has a Luttinger surface within the Brillouin zone. 
In the single-band model we are discussing, that may occur only at half-filling. Following Oshikawa~\cite{Oshikawa-PRL2000} we imagine to adiabatically thread in the above Mott insulator a fictitious flux quantum $\Phi_0$ that only couples to 
one spin species, whose particle number is conserved by charge $U(1)$ and spin $SU(2)$, assuming, e.g., a gauge in which the vector potential has only finite $x$-component. 
The final state differs from the initial one by a lattice momentum of $\pi$ in the $x$-direction~\cite{Oshikawa-PRL2000}, and that must be supplied by the 'quasiparticles' at the Luttinger surface~\cite{mio-2}. The same result holds true if we couple the flux to the other spin species. 
On the contrary, if the flux couples to both spin species, the system acquires a momentum $2\pi\equiv 0$, and that suggests 
that each spin species contributes with momentum $\pi$. The conclusion is that the Luttinger surface, whatever its shape and volume are, contributes to the particle count by one electron per site. If that remains true even when, upon doping 
the Mott insulator, Fermi pockets appear in the Brillouin zone, then Oshikawa's argument implies that the electron 
filling fraction $\nu$ is given by 
\beal
\nu &= \fract{1}{2} + v_{EP} - v_{HP}\,,\label{Oshikawa result}
\eal
where $v_{EP}$ and $v_{HP}$ are the fraction of electron-like and hole-like Fermi pockets with respect to the whole 
Brillouin zone. This result is consistent with the proposal of Yang, Rice and Zhang~\cite{Rice-PRB2006,Rice-RPP2011} in the pseudo-gap phase of underdoped cuprates,
but also of fractionalised Fermi liquids~\cite{Fractionalised-Senthil-PRL2003}. Equation~\eqn{Oshikawa result} is graphically shown in 
Fig.~\ref{count}.

\begin{figure}[htb]
\centerline{\includegraphics[width=0.4\textwidth]{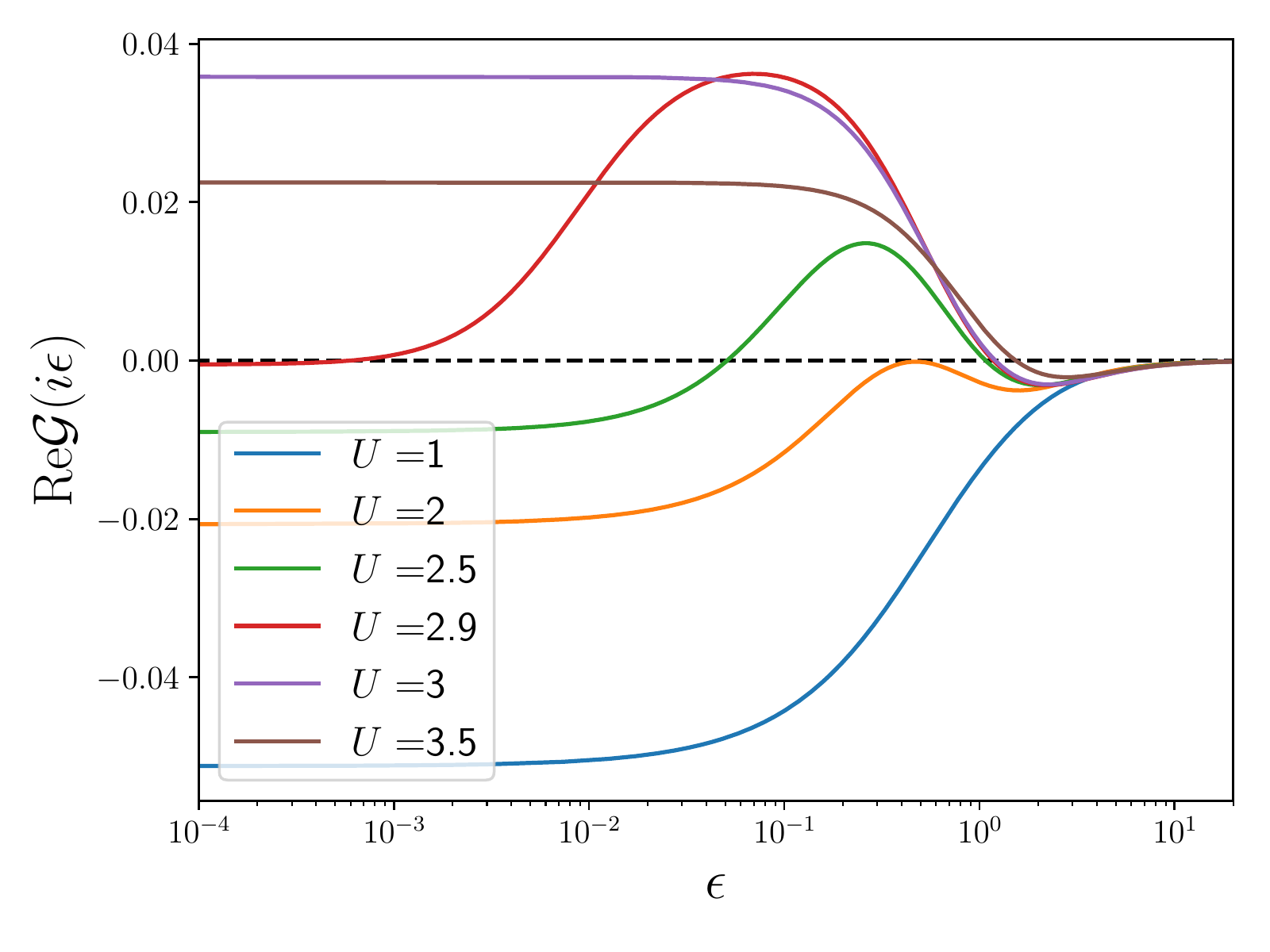}}
\caption{Real part of the local Green's function as obtained by dynamical mean field theory in the Hubbard model with 
a very weak chemical potential breaking particle-hole symmetry. Upon increasing the Hubbard $U$, the model has a transition between a metal and a Mott insulator, which, away from particle-hole symmetry, is first order. In the figure we show the evolution of $\Rea\mathcal{G}(i\ep)$ starting from the weak coupling metal and raising $U$. Note that a double zero first appear in the metal at $U\simeq 2$ at finite $\ep$, which signals the birth of the Hubbard bands. Upon further increasing $U$, that zero splits into two, one moving towards $\ep=0$. The value of $U\simeq 2.9$ at which the root 
reaches $\ep=0$ corresponds to the metal spinodal point, above which the only stable phase is insulating.}
\label{Rea G DMFT}
\end{figure}

To better understand how the situation depicted in Fig.~\ref{count} may occur, 
let us start from the perturbative regime and, upon varying the Hamiltonian parameters $\lambda$, like the interaction strength or the doping, reach the 
point $\lambda_c$ at which perturbation theory breaks down, i.e., its convergence radius. For convenience, we assume that $\lambda<\lambda_c$ identifies the perturbative regime, and $\lambda>\lambda_c$ the non-perturbative one. Therefore,
$\lambda=\lambda_c$ corresponds 
to the birth of a Luttinger surface and a concomitant 
dramatic change within the Brillouin zone: a large Fermi surface either disappears or abruptly turns into small hole and/or electron Fermi pockets, consistently with Eq.~\eqn{Oshikawa result}. Let us try to imagine how that may occur.
In general,  
$\Rea\,G(i\ep,\bk) =\Rea\,G(-i\ep,\bk)$ has an even number of roots $2\ell_\bk$ on the imaginary frequency axis, symmetrically located around $\ep=0$. If we borrow the results obtained in the Hubbard model by single-site dynamical mean field theory (DMFT)~\cite{DMFT-review}, see Fig.~\ref{Rea G DMFT}, and translate them in finite dimensions, we expect that at fixed $\ep=\ep_r>0$, which is function of $\lambda$ 
and vanishes as $\lambda\to \lambda_c$ from below, there 
is a surface of roots of $\Rea\,G\big(i\ep_r,\bk_L(\ep_r)\big)=\Rea\,G\big(-i\ep_r,\bk_L(\ep_r)\big)$, or, equivalently, of $\ep_*\big(\ep_r,\bk_L(\ep_r)\big)$, which smoothly evolves into the Luttinger surface as $\lambda\to\lambda_c$. 
Similarly, we can always define at any small $\ep$, thus also at $\ep_r$ when 
$\lambda\lesssim\lambda_c$, 
a surface of zeros of $\ep_*\big(\ep,\bk_F(\ep)\big)$ that are instead smoothly connected to the Fermi surface at $\ep=0$, i.e., the roots of $\ep_*(0,\bk_F)$. 
Since $\ep_*(\ep_r,\bk)$ are the eigenvalues of a hermitian operator, 
if the two surfaces, $\bk_L(\ep_r)$ and $\bk_F(\ep_r)$, cross within the Brillouin zone, those are actually avoided crossings. That simply rationalises the Fermi surface reshaping predicted by Eq.~\eqn{Oshikawa result}, see Fig.~\ref{count}, as $\lambda\to\lambda_c$, thus $\ep_r\to 0$.



In the case of Fig.~\ref{Rea G DMFT}, where the breakdown corresponds to the 
metal spinodal point, the two zeros at $\ep=\pm\ep_r$ simply annihilate each other
when $\ep_r\to 0$ as $\lambda\to\lambda_c$. Beyond single-site DMFT, we cannot exclude that the Luttinger surface survives after 
the breakdown, thus Eq.~\eqn{Oshikawa result}, 
changes shape and eventually disappears, as in the case discussed in Ref.~\cite{Rosh-2007}. Once that has happened, namely once the two zeros that had merged at $\ep=0$ finally annihilate each other, it is difficult to ascertain from the behaviour at $\ep=0$ whether 
the system is in the perturbative regime, and thus we can use 
conventional Luttinger's theorem, or, instead, perturbation theory has broken down and, in that case, how to count electron number.  
There is however a circumstance where we can 
make a firm statement, namely, when the self-energy is local, as in 
single-site dynamical mean field theory (DMFT)~\cite{DMFT-review}, see 
Fig.~\ref{Rea G DMFT}, or in impurity models.  In that case, the 
sign of the real part of the impurity Green's function $\mathcal{G}(i\ep)$, 
which is the local Green's function in DMFT, is fixed as $\ep\to\infty$, i.e., in the Hartree-Fock regime, and it is negative if the impurity is less than half-filled, the case of Fig.~\ref{Rea G DMFT}, and positive otherwise. It follows that, when perturbation theory is valid and Luttinger's theorem holds, 
then the sign of $\Rea\,\mathcal{G}(i\ep)$ at $\ep=0$ must be the same as at $\ep\to\infty$. When it breaks down, the sign must be opposite, corresponding to the two zeros 
of $\Rea\,\mathcal{G}(i\ep)$ that have annihilated each other at $\ep=0$. 
Therefore, the expectation value of the impurity occupation number close to half-filling is 
\beal
n &= \sum_\sigma  \Bigg(\,\fract{1}{2} - \int_0^\infty 
\fract{d\ep}{\pi}\,\fract{\partial \delta(\ep)}{\partial\ep}\Bigg)\\
&\qquad  -\sum_\sigma\,\fract{1-(-1)^{\ell}}{4}\,\text{sign}\big(\Rea\,\mathcal{G}(0)\big)\,,
\label{PT-NO-AIM}
\eal
where now $\delta(\ep)=\text{arg}\big(\mathcal{G}(i\ep)\big)$, and 
$\ell$ is simply the number of roots of $\Rea\,\mathcal{G}(i\ep)$ in the semi axis 
$0<\ep<\infty$. 

In what follows, we discuss few solvable cases where perturbation theory breaks down 
and Luttinger's theorem is violated, and test the validity of Eqs.~\eqn{Oshikawa result} or \eqn{PT-NO-AIM}.

\section{SDW fluctuation state}
\label{SDW}
The first example that we analyse is the model studied in Ref.~\cite{Altshuler-EPL1998} 
as representative of a nearly antiferromagnetic Fermi liquids. The model consists of electrons on a $V$-site cubic or square lattice, with non-interacting dispersion $\ep(\bk)$. The electrons exchange critical longitudinal spin fluctuations, with dynamical susceptibility  
\beal
\chi(i\,\omega,\bq) = \fract{\Delta^2}{g}\,\fract{\delta_{\omega,0}}{T}\,V\,\delta_{\bq,\bQ}\,,
\label{SDW-U}
\eal
where $\bQ=(\pi,\dots,\pi)$ and $g$ is the exchange constant. The exact self-energy in the paramagnetic phase reads~\cite{Altshuler-EPL1998,Chubukov-PhsRep1997} 
\be
\Sigma(i\ep,\bk) = \fract{\Delta^2}{\;i\ep-\ep(\bk+\bQ)\;} = 
\Delta^2\,G_0(i\ep,\bk+\bQ)\;,
\ee
where $G_0(i\ep,\bk)$ is the non-interacting Green's function, 
hence
\beal
G^{-1}(i\ep,\bk) &= G_0^{-1}(i\ep,\bk) -\Delta^2\,G_0(i\ep,\bk+\bQ)\,.\label{SDW-G}
\eal

\begin{figure}
\centerline{\includegraphics[width=0.48\textwidth]{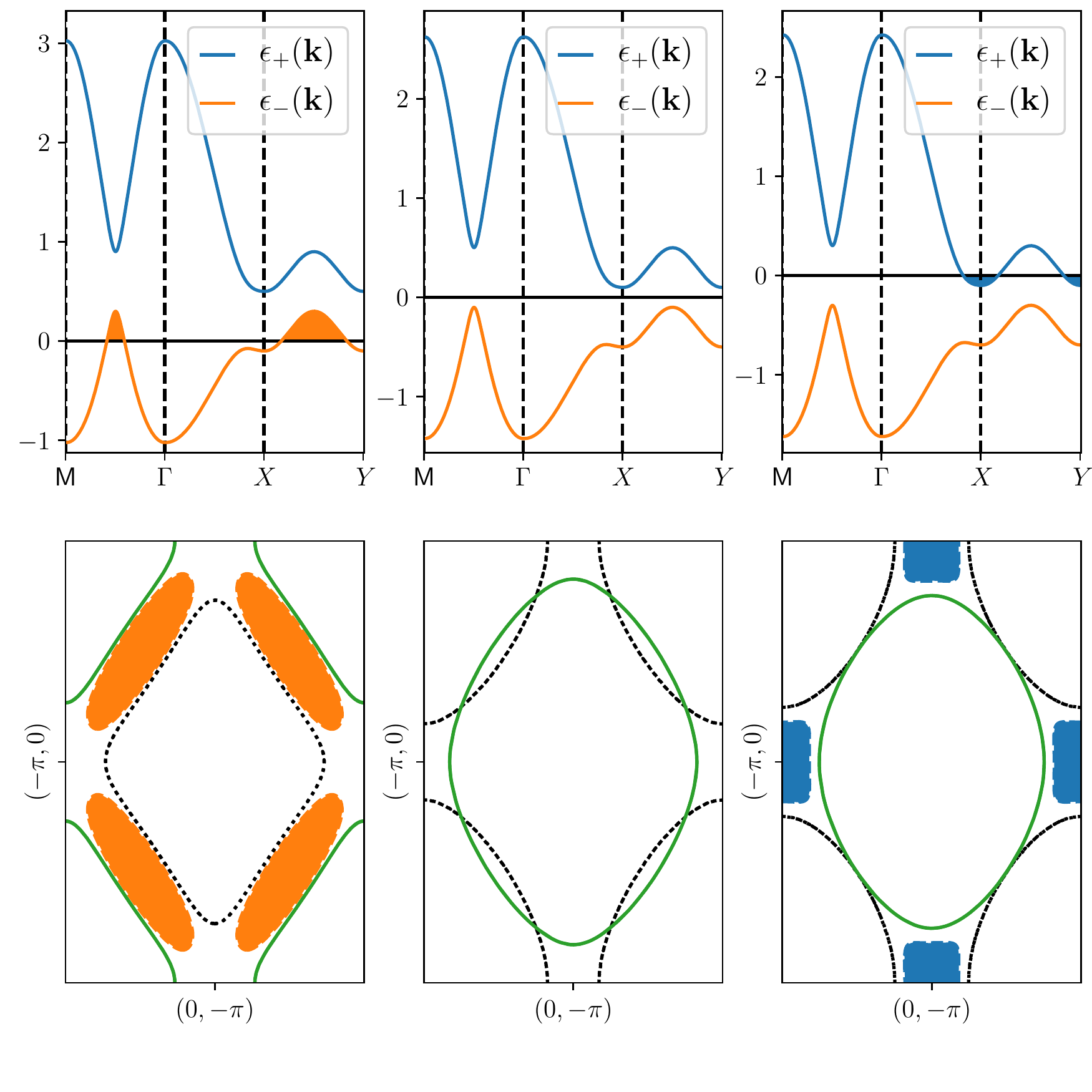}}
\caption{Top panels: band structure Eq.~\eqn{SDW:ep+ep-} on a square lattice 
with nearest $t$ and next nearest, $t'=-0.2t$, neighbour hopping, at $\Delta=-0.3$ 
and different chemical potentials corresponding to hole, left panel, and electron, right panel, doping with respect to half-filling, middle panel.
Bottom panels: corresponding Luttinger surface, green line, and Fermi pockets, hole-like in orange and electron-like in blue. The non-interacting Fermi surface is also shown, black dotted line. In the present case, our conjecture \eqn{Oshikawa result}
predicts that all $\bk$ points in the Brillouin zone contribute with one to the total electron number, with the exception of  those inside the 
Fermi pockets, which contribute with zero or with two if the pockets are, respectively, hole- or electron-like.}
\label{FermiPockets}
\end{figure}

In this case a Luttinger surface always exists and  Luttinger's theorem is violated at any $\Delta\not=0$~\cite{Altshuler-EPL1998}. Through the exact Green's function \eqn{SDW-G} one readily finds~\cite{Altshuler-EPL1998} that, for a single spin species,  
\beal
n(\bk)+n(\bk+\bQ) = \theta\big(\ep_+(\bk)\big) + \theta\big(\ep_-(\bk)\big)\,,
\label{SDW-exact}
\eal
where 
\beal
\ep_{\pm}(\bk) &= \fract{\ep(\bk)+\ep(\bk+\bQ)}{2} \\
&\qquad \pm \sqrt{\left(\fract{\ep(\bk)-\ep(\bk+\bQ)}{2}\right)^2 + \Delta^2\,}\;,
\label{SDW:ep+ep-}
\eal
see Fig.~\ref{FermiPockets}, so that 
\begin{itemize}
\item $n(\bk)+n(\bk+\bQ)= 2\theta\big(-\ep(\bk)\big)$ if 
$\ep(\bk)\ep(\bk+\bQ) >\Delta^2>0$,
\item $n(\bk)+n(\bk+\bQ)=1$ if   
$\Delta^2>\ep(\bk)\ep(\bk+\bQ)$.
\end{itemize}
The quasiparticle residue at $\ep=0$ is now 
\beal
Z(\bk) = \fract{\ep(\bk+\bQ)^2}{\Delta^2+\ep(\bk+\bQ)^2}\;,
\eal
so that the Luttinger surface is defined by $\bk_L:\,Z(\bk_L)=0$, i.e., 
$\bk_L:\,\ep(\bk_L+\bQ)=0$, while the quasiparticle energy 
by 
\beal
\ep_*(\bk) = Z(\bk)\,\fract{1}{\ep(\bk+\bQ)}\,
\Big(\ep(\bk)\,\ep(\bk+\bQ)-\Delta^2\Big)\,,
\label{SDW: quasiparticle energy}
\eal
which allows defining the Fermi surface by $\bk_F:\,\ep(\bk_F)\,\ep(\bk_F+\bQ)=\Delta^2$. 
The non-interacting Fermi surface, the interacting Luttinger one, and the interacting Fermi pockets are shown in Fig.~\ref{FermiPockets} for few exemplary cases. 
Let us now apply Eq.~\eqn{Oshikawa result} to calculate the momentum distribution. Through $\ep_*(\bk)$ 
in Eq.~\eqn{SDW: quasiparticle energy} we realise that the Fermi pockets, when they exist, include all 
$\bk$ such that $\ep(\bk)\,\ep(\bk+\bQ)\geq \Delta^2$, and are electron-like if $\ep(\bk)<0$ and hole-like otherwise. 
This observation together with Eq.~\eqn{Oshikawa result} directly yields Eq.~\eqn{SDW-exact}. 

Despite its simplicity, this model is very instructive and yields insights that 
we believe are rather general. Since the interaction is a $\delta$-function in frequency, it is rather easy to express the 
self-energy as functional of the interacting Green's functions \eqn{SDW-G} and 
of the interaction strength $\Delta$. We find that
\be
\Sigma(i\ep,\bk) = \fract{\;\sqrt{1+X(i\ep,\bk,\bk+\bQ)\;}-1\;}{2G(i\ep,\bk)}\,,\label{SDW:Sigma.vs.G}
\ee
where
\beal
X(i\ep,\bk,\bk+\bQ) \equiv 4\Delta^2\,G(i\ep,\bk)\,G(i\ep,\bk+\bQ)\,,\label{SDW-X}
\eal
through which the Luttinger integral can be written as 
\bw
\beal
I_L(\bk,\bk+\bQ) &= \int_{-\infty}^\infty \fract{d\ep}{2\pi}\;\Bigg\{
G(i\ep,\bk)\fract{\partial\Sigma(i\ep,\bk)}{\partial i\ep}
+G(i\ep,\bk+\bQ)\fract{\partial\Sigma(i\ep,\bk+\bQ)}{\partial i\ep}\Bigg\}\\
&= \int_{-\infty}^\infty \fract{d\ep}{2\pi}\;\fract{\partial}{\partial i\ep}\,\ln\fract{\sqrt{1+X(i\ep,\bk,\bk+\bQ)\,}+1}{2}\\
&= -\fract{1}{\pi}\;\Ima\,\ln\fract{\sqrt{1+X(i0^+,\bk,\bk+\bQ)\,}+1}{2}\\
&= -\theta\big(\Delta^2-\ep(\bk)\,\ep(\bk+\bQ)\big)\,
\theta\big(\ep(\bk)\,\ep(\bk+\bQ)\big)\,\text{sign}\big(\ep(\bk)+\ep(\bk+\bQ)\big)
\,,\label{nota-bene}
\eal
\ew
consistently with Eq.~\eqn{I_L-deriv}.
It is worth noticing that $I_L$ yields an entanglement between the phases 
$\delta(0,\bk)$ and $\delta(0,\bk+\bQ)$ of the two Green's functions, which 
appear as independent quantities in conventional Luttinger's theorem. We believe that is the key role of the Luttinger integral whenever it is finite.\\ 
We can take a step further and explicitly built the Luttinger-Ward functional 
\be
\Phi[G] = \sum_{\bk}T\sum_n\esp{i\ep_n\eta}\, \Phi\big[G(i\ep_n,\bk),G(i\ep_n,\bk+\bQ)\big]\,,
\ee
where the sum over $\bk$ is within the reduced Brillouin zone, by solving  
\beal
\fract{\delta \Phi\big[G(i\ep,\bk),G(i\ep,\bk+\bQ)\big]}{\delta G(i\ep,\bk)}
= \Sigma(i\ep,\bk)\,.
\eal
We find that $\Phi\big[G(i\ep_n,\bk),G(i\ep_n,\bk+\bQ)\big]$ is actually 
a functional $\Phi[X]$ of $X$ in Eq.~\eqn{SDW-X}, specifically
\beal
\Phi[X] &= \sqrt{1+X\,}-1 - \ln\fract{\sqrt{1+X\,}+1}{2}\;.
\eal
We end noticing that the square root 
in the expression \eqn{SDW:Sigma.vs.G} of $\Sigma(i\ep,\bk)$  
implies that the inverse of Dyson's equation 
\beal
G_0(i\ep,\bk)^{-1} &= G(i\ep,\bk)^{-1} + \fract{\delta\Phi[G]}{\delta G(i\ep,\bk)}\;,
\eal
generally admits two solutions $G_0(i\ep,\bk)$, only one of which is physical. 
This result agrees with several evidences~\cite{Georges-PRL2015,Schafer-PRB2016,Toschi-PRL2017,Chalupa-PRB2018} that the Luttinger-Ward functional may become multivalued upon increasing the interaction strength. 

\section{Pseudo-gap impurity model}
\label{AIM}
\begin{figure}
\vspace{-0.5cm}
\centerline{\includegraphics[width=0.45\textwidth]{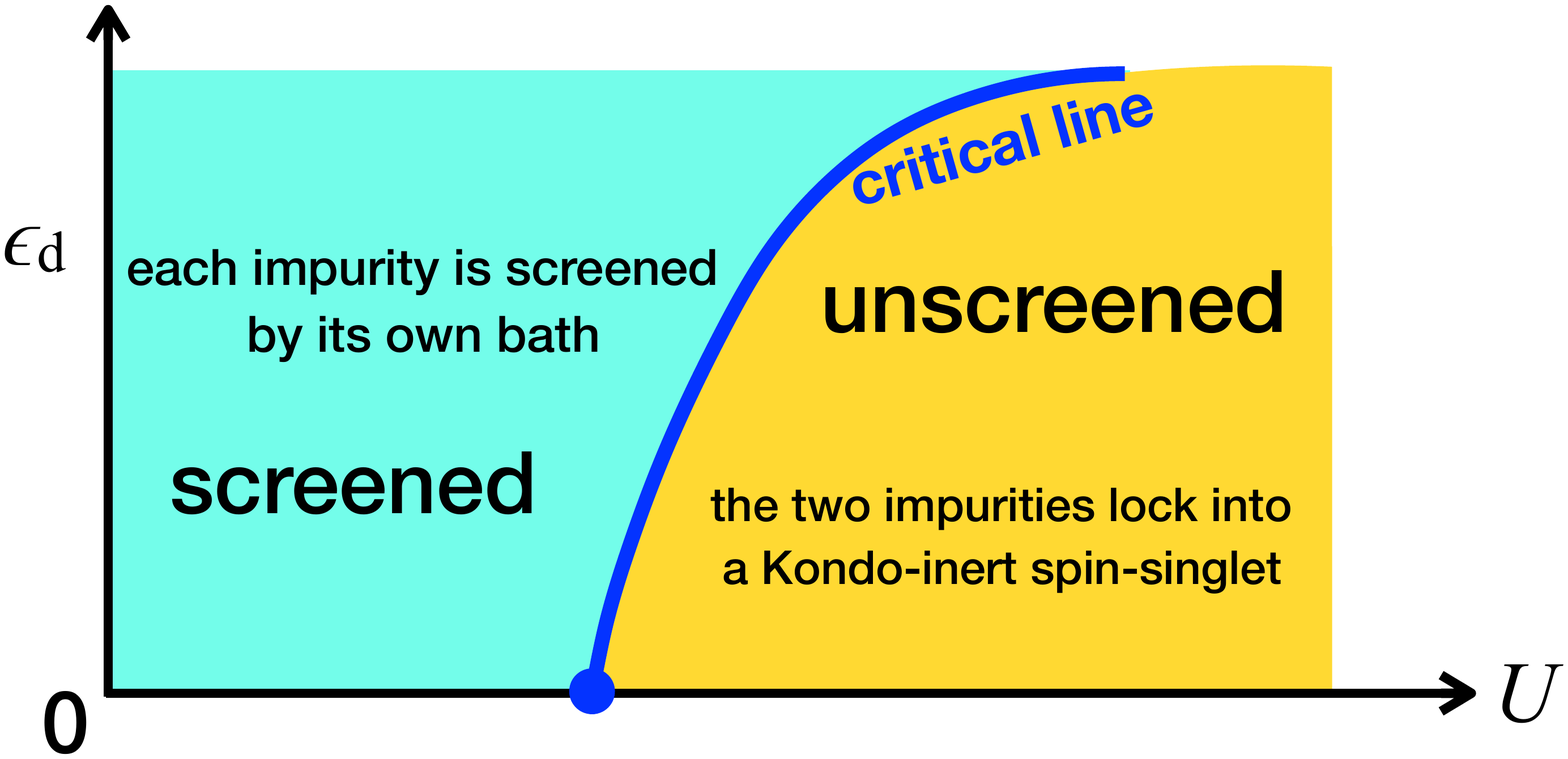}}
\vspace{-0.5cm}
\caption{Phase diagram of the impurity model \eqn{HAIM} at fixed hybridisation width $\Gamma$ and 
$J\ll \Gamma$, as function of $U$ and $\ep_d>0$. The case $\ep_d<0$ is symmetric. }
\label{phase diagram}
\end{figure}

Let us now discuss the failure of Luttinger's theorem in the impurity model studied in Ref.~\cite{Hewson-2018} by numerical renormalisation group (NRG). For convenience, we consider a slightly different model with the same physical properties, which was thoroughly investigated in Ref.~\cite{Lorenzo-PRB2004} thus saving us from 
recalculating the whole phase diagram. The model represents a two-orbital Anderson impurity with 
inverted Hund's rules. The Hamiltonian is 
\beal
H &=H_0+H_\text{imp}\,,\label{HAIM}
\eal
where 
\be
H_0 = \sum_{i=1}^2\sum_{\bk\sigma}\Big[\ep_\bk\,c^\dagger_{i\bk\sigma}\,c^\dagga_{i\bk\sigma} 
+ V_\bk\,\big(c^\dagger_{i\bk\sigma}\,d^\dagga_{i\sigma} + H.c.\big)\Big]\,,\label{H0AIM}
\ee
is the sum of two equivalent resonant level models, and 
\be
H_\text{imp} = \ep_d\,\big(n-2\big)
+ \fract{U}{2}\big(n-2\big)^2-2J\,\Big( \bd{T}\cdot\bd{T} - T_3^2\Big)\,,\label{Hlocal}
\ee
where $n=\sum_{i\sigma} n_{i\sigma}$, with  $n_{i\sigma}=d^\dagger_{i\sigma}\,d^\dagga_{i\sigma}$ the occupation number of the impurity orbital $i=1,2$ with spin $\sigma$, while $\bd{T} = \big(T_1,T_2,T_3\big)$ is a pseudo-spin 
operator with 
\beal
T_a &=\fract{1}{2}\,\sum_\sigma\,\sum_{ij}\, 
d^\dagger_{i\sigma}\,\big(\hat{\tau}_a\big)_{ij}\,d^\dagga_{j\sigma}\,,& a=1,2,3\,,
\eal
and $\hat{\tau}_a$ the Pauli matrices in the two-orbital space. We assume that $H_0$ in \eqn{H0AIM} 
is particle-hole (p-h) symmetric, so that a finite $\ep_d$ in \eqn{Hlocal} 
is the only source of p-h symmetry breaking. In the following calculations we take 
a hybridisation width  
\beal
\Gamma(\ep) &\equiv \pi\,\sum_\bk\,V_\bk^2\,\delta\big(\ep-\ep_\bk\big)
= \Gamma\,\theta\big(1-|\ep|\big)\,, 
\eal
with $\Gamma=0.1$, which also defines our unit of energy, and $J=0.004\ll\Gamma$.\\
When $U$ is large, the impurity is occupied by two electrons that can form a spin-triplet orbital-singlet
($S=1$, $T=0$), or a spin-singlet orbital-triplet ($S=0$, $T=1$). If $J>0$, as we assume, the lowest 
energy state 
\beal
\fract{1}{\sqrt{2}}\,\big(d^\dagger_{1\up}\,d^\dagga_{2\down}+d^\dagger_{2\up}\,d^\dagga_{1\down}
\big)\ket{0}\,,\label{0}
\eal
has $S=0$, $T=1$, and $T_3=0$. If we regard the two orbitals as the single orbitals of two impurities, state 
\eqn{0} simply represents the two impurities coupled into a spin-singlet configuration. In other words, for large $U$ the Hamiltonian \eqn{HAIM} 
is actually equivalent to two spin-1/2 impurities, each Kondo coupled to its own bath, and coupled to each 
other by an antiferromagnetic exchange, which is the model studied in Ref.~\cite{Hewson-2018}. The phase diagram of this model
depends on the magnitude of $J$ relative to the Kondo temperature 
$T_K$ at $J=0$. If $J\ll T_K$, each impurity is Kondo screened by its bath, leading to 
a conventional Kondo effect. On the contrary, if $J\gg T_K$, the two impurities lock into a Kondo-inert spin-singlet state. These two regimes, which we denote as 'screened' and 'unscreened' phases, are separated by a 
quantum critical point~\cite{Jones&Varma,*Jones&VarmaPRB}, actually a whole critical line at $\ep_d\not=0$~\cite{Lorenzo-PRB2004}. Since we work at constant $\Gamma$ and 
$J\ll \Gamma$, and $T_K$ decreases with increasing $U$, the critical point is reached upon increasing $U$. Specifically, with the chosen $\Gamma$ and $J$, its location is at  $U_c\simeq 1.85$ when $\ep_d=0$. In Fig.~\ref{phase diagram} 
we sketch the phase diagram as function of $U$ and $\ep_d>0$.\\
The screened, $U<U_c$, and unscreened, $U>U_c$, phases are both local Fermi liquids in Nozi\`eres sense~\cite{Nozieres-localFL}, despite the unscreened phase is not adiabatically connected to the non-interacting limit $U=J=0$. 
For instance, at p-h symmetry, $\ep_d=0$, the impurity self-energy in the unscreened phase diverges at the Fermi level~\cite{Lorenzo-PRB2004}, the local counterpart of a 
Luttinger surface, which leads to a pseudo gap 
in the density-of-states that is gradually filled in when $\ep_d\not= 0$~\cite{Lorenzo-PRB2004}; a totally different behaviour from a non-interacting resonant level model.

\subsection{Fate of Luttinger's theorem in the impurity model}

\begin{figure}[ht]
\vspace{-0.5cm}
\centerline{\includegraphics[width=0.49\textwidth]{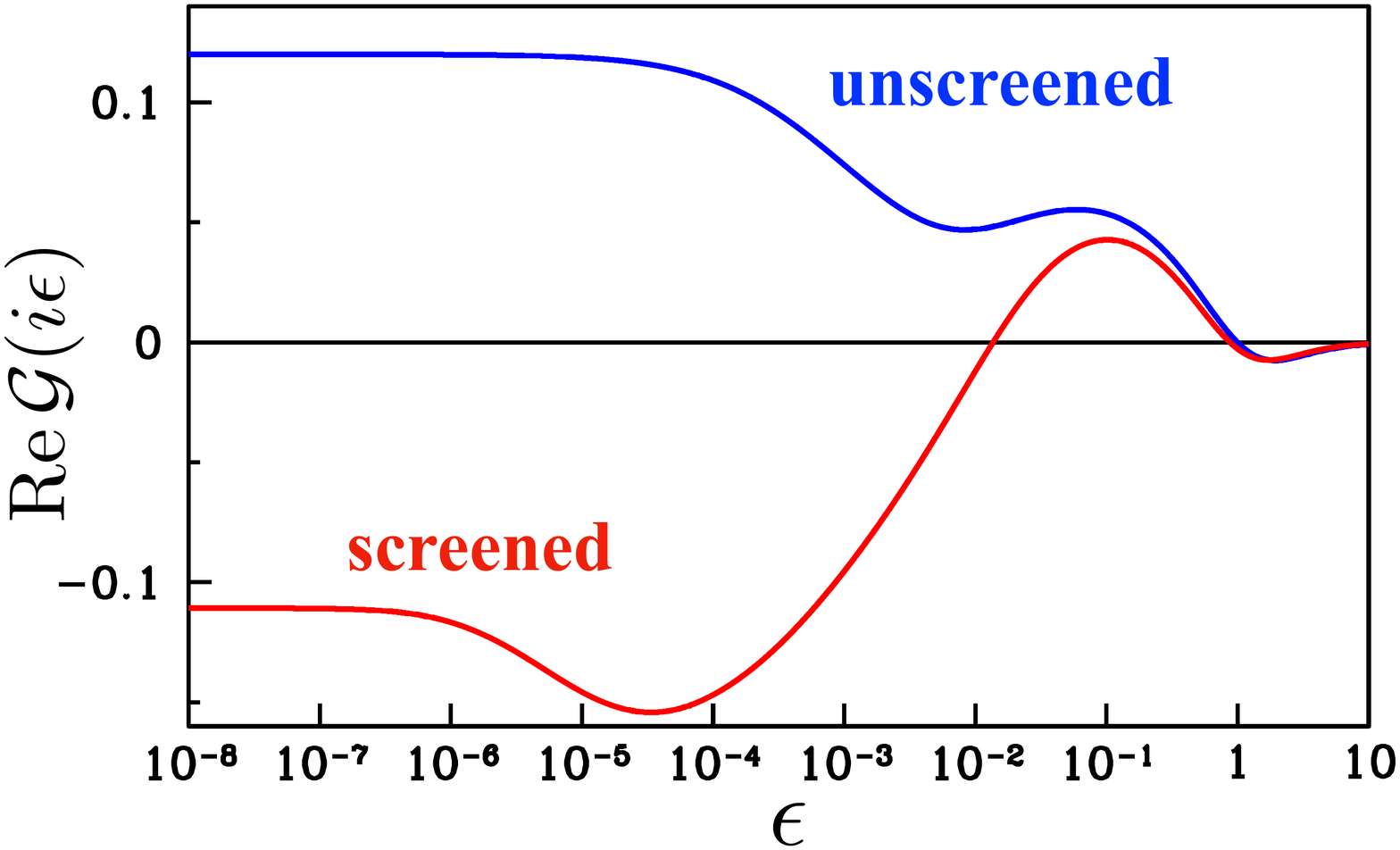}}
\vspace{-0.8cm}
\caption{$\Rea\,\mathcal{G}(i\ep)$ versus the Matsubara frequency $\ep$ at 
$\ep_d=0.1$ in the screened phase, $U=1.75$ red curve, and in the unscreened one, $U=2$ blue curve. }
\label{ReG}
\end{figure}

The Hamiltonian \eqn{HAIM} at $\ep_d\not=0$ is invariant under global spin $SU(2)$,  separate charge $U(1)$ rotations in each channel $i=1,2$, that includes the conduction bath and the corresponding 
impurity level, as well as under the $Z_2$ symmetry $1\leftrightarrow 2$. If the conduction bandwidth is large enough, as we assume hereafter, the conserved quantities become effectively those at the impurity site, since the fluctuations in the bath are negligible. The impurity Green's function is 
\be
\mathcal{G}(i\ep) = \fract{1}{i\ep -\ep_d +i\,\Gamma-\Rea\,\Sigma(i\ep) -i\,\Ima\,\Sigma(i\ep)}\;,
\ee
and, by symmetry, is independent of $i=1,2$ and $\sigma=\up,\down$, and therefore Eq.~\eqn{PT-NO-AIM} 
reads 
\beal
n_{i\sigma} &= \left(\fract{1}{2} - \fract{
\text{arg}\big(\mathcal{G}(i\infty)\big)-
\text{arg}\big(\mathcal{G}(0)\big)}{ \pi}\right)\\
&\qquad  - \fract{1-(-1)^\ell}{4}\;
\text{sign}\big(\Rea\,\mathcal{G}(0)\big) \;,\label{n-2-AIM}
\eal
where the term in parentheses is just the conventional statement of Luttinger's 
theorem that was shown in Ref.~\cite{Hewson-2018} not to yield the 
correct result in the unscreened phase at $\ep_d\not=0$. 
The last term in Eq.~\eqn{n-2-AIM}, which corrects that result 
when Luttinger's theorem fails, is finite only when the number $\ell$ of 
zeros of $\Rea\,\mathcal{G}(i\ep)$ for $0<\ep<\infty$ is odd.
Fig.~\ref{ReG} shows $\Rea\,\mathcal{G}(i\ep)$ in the screened and unscreened phases at $\ep_d=0.1$. Not surprisingly, $\ell$ is even in the screened phase, 
and odd in the unscreened one, in which  
case the last term in Eq.~\eqn{n-2-AIM} 
is finite and equal to $-1/2$. For $\ep_d<0$ the correction is actually $+1/2$ since the real part of $\mathcal{G}(i\ep)$ changes signs after a particle-hole transformation that brings $\ep_d\to-\ep_d$.\\
The $\pm 1/2$ correction is exactly the missing quantised term noticed in Ref.~\cite{Hewson-2018}, and thus 
Eq.~\eqn{n-2-AIM} does reproduce the correct electron number. 
We note that Fig.~\ref{ReG} explicitly demonstrates that, crossing the point at which perturbation theory breaks down, 
$\ell$ changes by one, from $\ell=2$ in the screened 
phase to $\ell=1$ in the unscreened one, as earlier discussed.

Besides the electron number, $n=\sum_{i\sigma}\,n_{i\sigma}$, the Hamiltonian \eqn{HAIM} admits other conserved quantities, e.g., the magnetisation 
$m=\sum_{i}\,(n_{i\up}-n_{i\down})$ and the relative orbital occupancy 
$n_f=\sum_{\sigma}\,(n_{1\sigma}-n_{2\sigma})$. A field that couples to any of 
those conserved quantities does not spoil the quantum critical point~\cite{Lorenzo-PRB2004}. We may then wonder whether conventional Luttinger's theorem also fails in providing the values of those quantities as it 
does for the electron number when crossing the critical point. Let us consider, for instance, the magnetisation $m$. According to Luttinger's theorem, we could calculate $m$ through 
\beal
m_L = \sum_{i=1}^2\,
\fract{\text{arg}\big(\mathcal{G}_{i\up}(0)\big)-
\text{arg}\big(\mathcal{G}_{i\down}(0)\big)
}{ \pi}\;.\label{m_L}
\eal
Evidently,  both $m$ and $m_L$ vanish when $SU(2)$ symmetry holds. Therefore, we add to the Hamiltonian \eqn{HAIM} with $\ep_d=0.1$ a Zeeman splitting term $-B\,m$, 
with very small $B=0.0001$ that nonetheless makes $\mathcal{G}_{1\up}(i\ep)= \mathcal{G}_{2\up}(i\ep)\not = \mathcal{G}_{1\down}(i\ep)= \mathcal{G}_{2\down}(i\ep)$. 
\begin{figure}[ht]
\centerline{\includegraphics[width=0.49\textwidth]{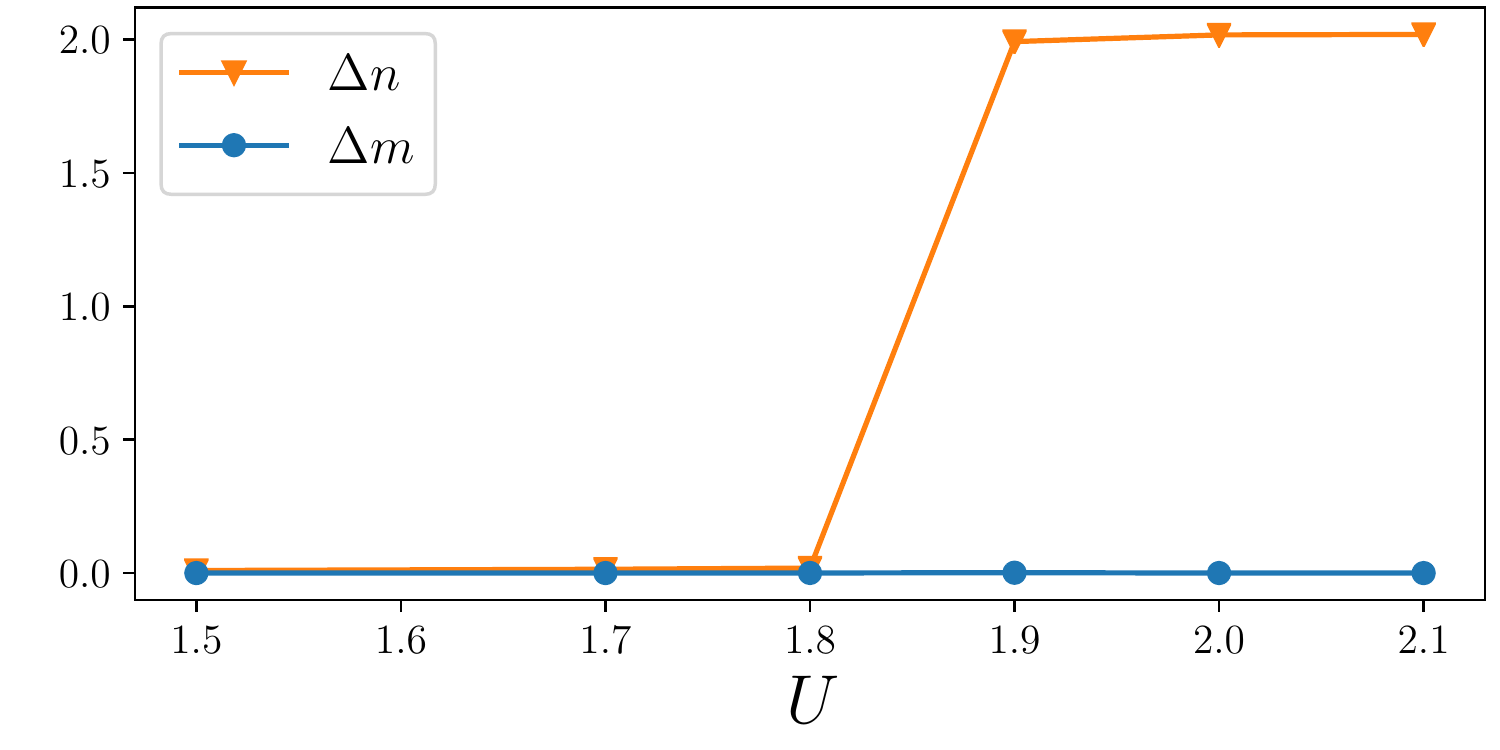}}
\caption{The behaviour of $\Delta n=n_L-n$ and $\Delta m=m_L-m$ as function of $U$ at $\ep_d=0.1$ and $B=0.0001$. Here, $n_L$ and $m_L$ are, respectively, the electron number and magnetisation calculated through Luttinger's theorem, while $n$ and $m$ their actual value.}
\label{magnetic}
\end{figure} 
In Fig.~\ref{magnetic} we show the deviation $\Delta m$ of $m_L$ in Eq.~\eqn{m_L} 
from the actual value $m$ as function of $U$. For comparison, we also plot the deviation $\Delta n$ of the 
Luttinger's theorem prediction for the number of particles,
\beal
n_L &= 4 + \sum_{i\sigma}\,\fract{\;\text{arg}\big(\mathcal{G}_{i\sigma}(0)\big)\;}
{\pi}\;,\label{n_L}
\eal
from the correct result $n$. We observe that while 
$\Delta n$ jumps from 0 to 2 crossing the critical point, consistent with the missing contribution from the Luttinger integral, see Eq.~\eqn{n-2-AIM}, $\Delta m$ remains always zero, showing that the corresponding 
Luttinger integral vanishes also in the unscreened phase, despite the breakdown of perturbation theory. In reality, if we instead take $B\gg\ep_d$, the situation is reversed: $\Delta m$ jumps from 0 to -2, while $\Delta n$ remains zero. \\
More generally, if we add different fields $\ep_d$, $B$ and $B_f$ that couple to $n$, $m$ and $n_f$, respectively, 
the strongest one identifies the channel where Luttinger's theorem breaks down, whereas the theorem still applies for the other two channels.

\subsection{The unscreened phase as paradigm of a pseudo-gapped metal}
\label{The unscreened phase as paradigm of a pseudo-gapped metal}
\begin{figure}[ht]
\centerline{\includegraphics[width=0.49\textwidth]{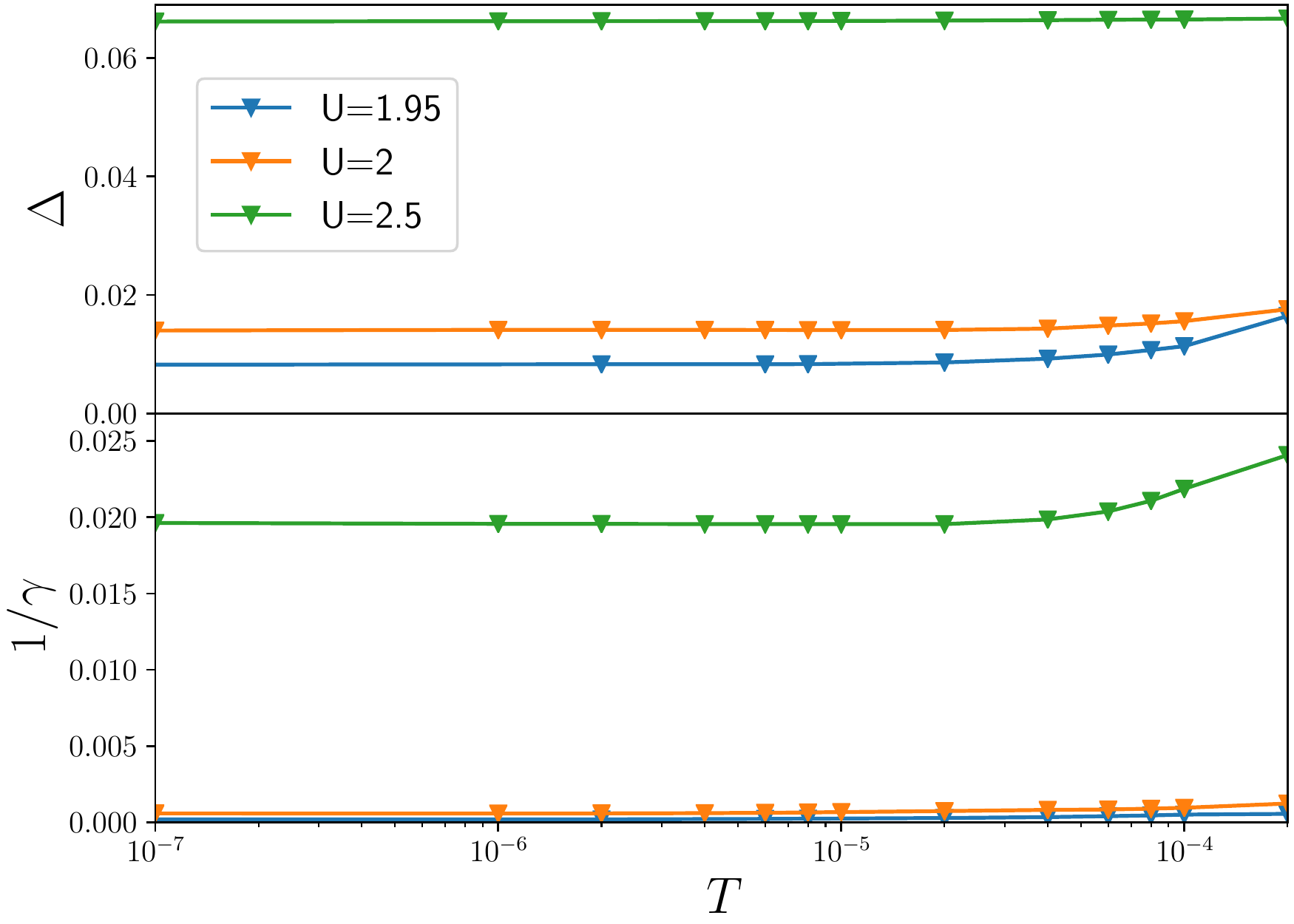}}
\vspace{-0.4cm}
\caption{Temperature dependence of the parameters $\Delta$ and $\gamma$ in Eq.~\eqn{fit} fitted through NRG results  at $\ep_d=0.1$ and different $U>U_c$. We note that both $\Delta^2$ and $1/\gamma$ vanish approaching 
$U_c$, although $1/\gamma \ll \Delta^2$.}
\label{fit-par}
\end{figure}

Lot of effort has been put over the past decades into modelling the self-energy of the pseudo-gap phase in underdoped cuprates~\cite{Rice-PRB2006,Rice-RPP2011,Imada-PRL2011,Alexei-RPP2019}, 
also revealed by cluster extensions of dynamical mean field theory in 
the Hubbard model doped away from the half-filled Mott insulator 
\cite{Kotliar-PRB2006,Civelli-PRL2016,Georges-PNAS2018,Georges-PRX2018}. 
Since the unscreened phase of the impurity model \eqn{HAIM} 
is also pseudo-gapped~\cite{Lorenzo-PRB2004}, it is worth modelling its self-energy, which is easily accessible by NRG at and away from p-h symmetry, as well as at zero and finite temperature. \\
\begin{figure}[ht]
\centerline{\includegraphics[width=0.49\textwidth]{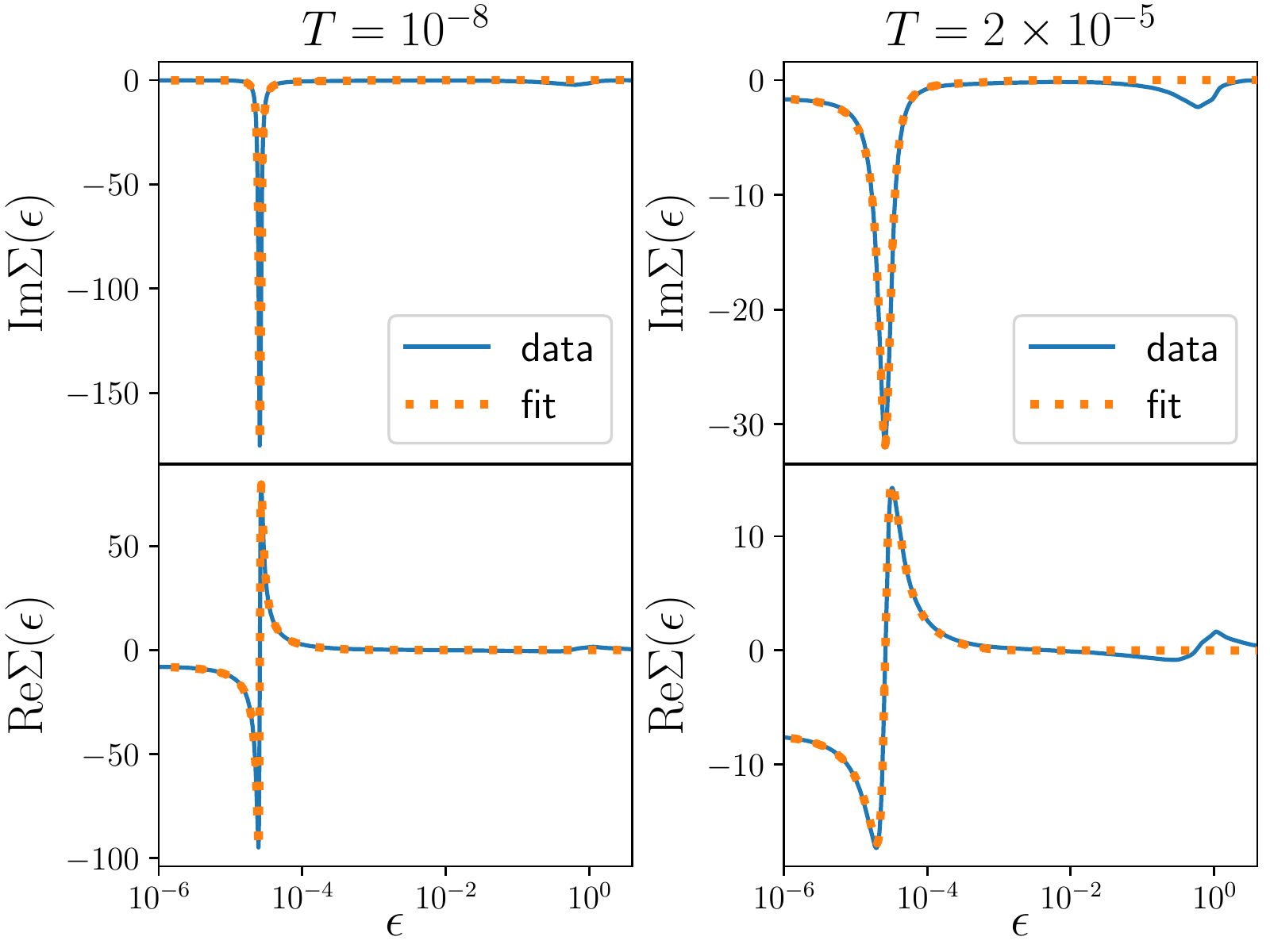}}
\caption{Real and imaginary parts of the retarded self-energy at $U=2$, $\ep_d=0.1$ 
and temperatures $T=10^{-8}$, left panels, and $T=2\times 10^{-5}$, right panels, 
together with the ansatz \eqn{fit} with fitted parameters (dotted lines).}
\label{fit-Sigma}
\end{figure} 
We find that the retarded impurity self-energy $\Sigma_+(\ep) \equiv \Sigma(\ep+i0^+)$ in the unscreened phase is well fitted at low energy $\ep$ and temperature $T$ by~\cite{mio,mio-2}, see Fig.~\ref{fit-Sigma}, 
\beal
\Sigma_+(\ep) = \fract{\Delta^2}{\;\ep - \mu + i\,\gamma\,\big(\ep^2+\pi^2\,T^2\big)\;}\;,
\label{fit}
\eal
where all real parameters $\Delta$, $\mu$ and $\gamma$ depend on $U$, $T$ and on the strength $\ep_d$ of the p-h symmetry breaking term. In particular, $\Delta^2$ and $1/\gamma$ vanish quadratically approaching the critical line $U=U_c$~\cite{Lorenzo-PRB2004}, while, consistently with Fig.~\ref{ReG}, $\mu$ has the same sign of $\ep_d$ and vanishes at $\ep_d=0$. In Fig.~\ref{fit-par} we show the parameters $\Delta$ and $\gamma$ extracted by the fit as function of $T$ and different $U>U_c$ at $\ep_d=0.1$. \\

\begin{figure}[hbt]
\centerline{\includegraphics[width=0.4\textwidth]{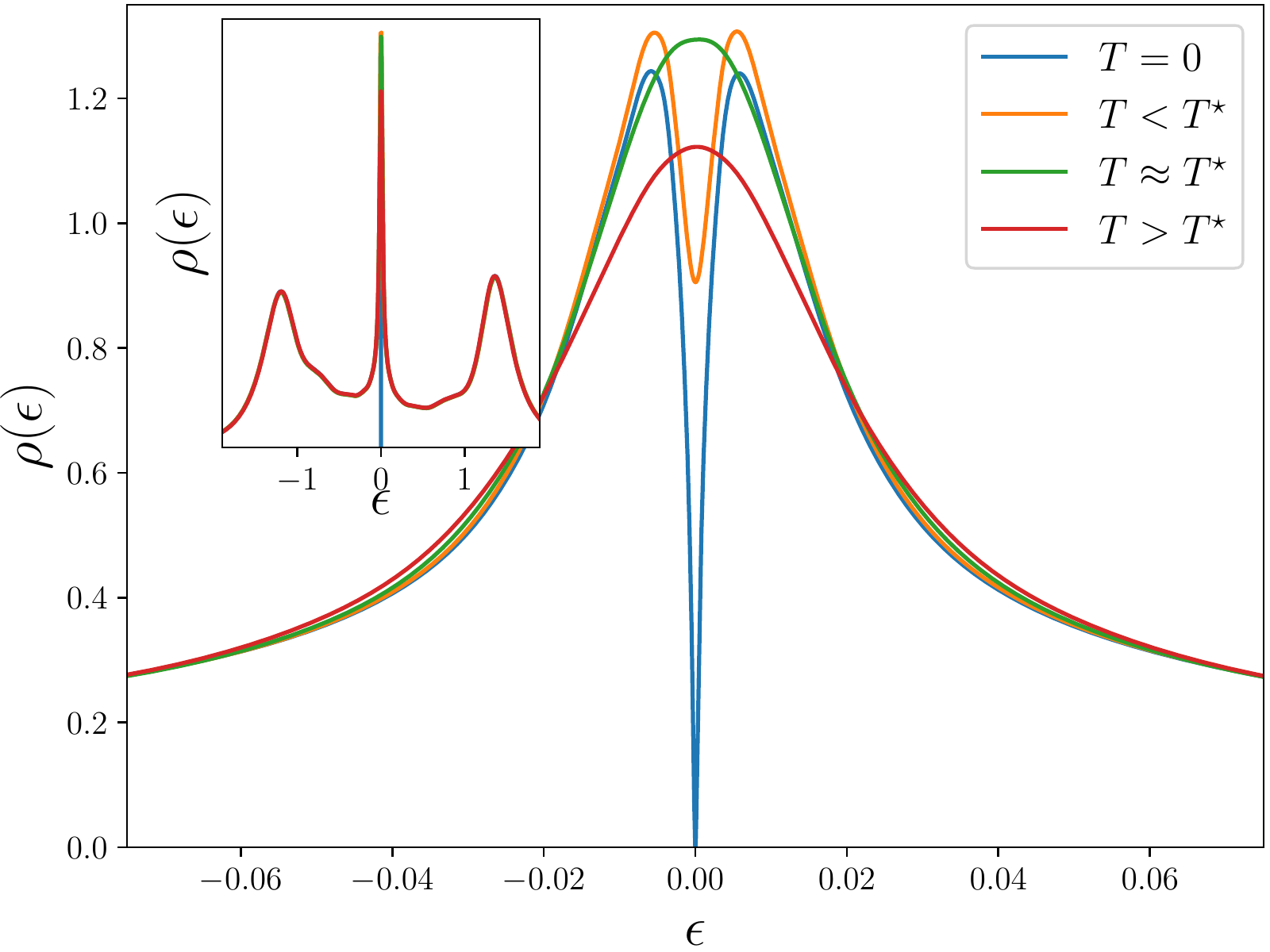}}
\caption{Impurity density of states at $U=2$, $\ep_d=0.1$ and different temperatures 
below and above the pseudogap temperature $T_*$.}
\label{DOS.vs.T}
\end{figure}

\begin{figure}[hbt]
\centerline{\includegraphics[width=0.4\textwidth]{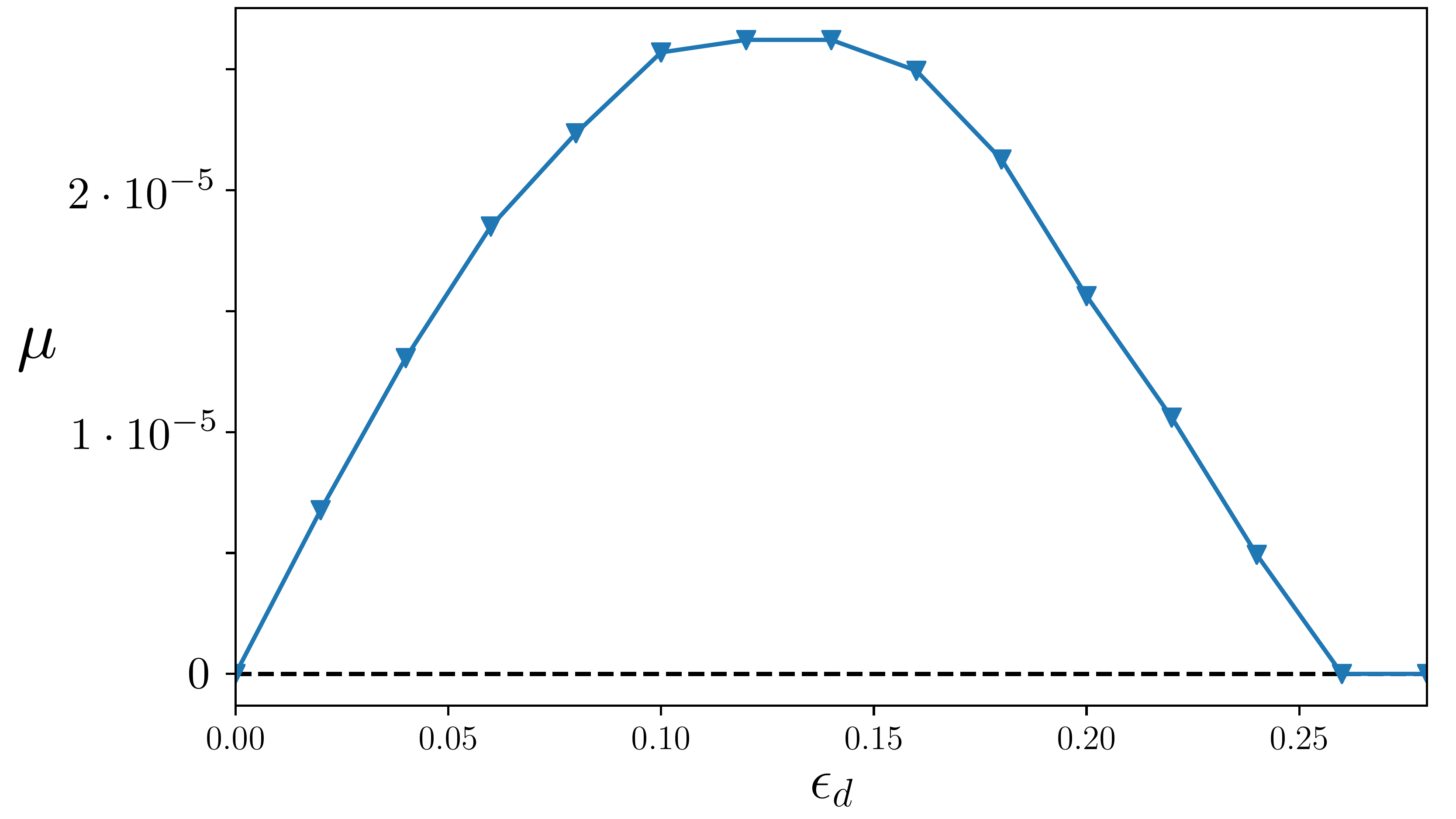}}
\caption{Parameter $\mu$ in Eq.~\eqn{fit} at $T=10^{-8}$ and $U=2$ 
as function of $\ep_d$ from 0 
to above the critical point, see Fig.~\ref{phase diagram}.}
\label{mu T}
\end{figure}

\noindent
At $T=\ep_d=0$, 
\beal
\Sigma_+(\ep) = \fract{\Delta^2}{\;\ep  + i\,\gamma\,\ep^2\;}\underset{\ep\to 0}{\simeq}
\fract{\Delta^2}{\;\ep\;} - i\,\gamma\,\Delta^2\,,\label{Sigma-ph}
\eal
corresponds to the highly singular expression found in Ref.~\cite{Lorenzo-PRB2004}, which, as earlier mentioned, is the local counterpart of a Luttinger surface. On the contrary, at $\ep_d\not= 0$ 
and for $\ep,T \ll \mu$,  
\be
\Sigma_+(\ep) \simeq -\fract{\Delta^2}{\mu} - \fract{\Delta^2}{\mu^2}\;\ep
-i\,\fract{\Delta^2\,\gamma}{\mu^2}\;\big(\ep^2+\pi^2\,T^2\big)\,,\label{Sigma-no-ph}
\ee
has a conventional Fermi-liquid behaviour, despite the spectral function pseudo-gap, and the `Luttinger surface` has disappeared.\\
The quantum critical point entails the existence at finite temperature of a quantum critical region delimited by a crossover temperature $T_*$ that, in the unscreened 
phase, can be identified with the temperature below which the pseudo gap opens, see  
Fig.~\ref{DOS.vs.T} where we plot the impurity density-of-states (DOS) $\rho(\ep)$, 
at $\ep_d=0.1$, $U=2$ and different $T$.
The phase diagram~\ref{phase diagram} shows that the critical line can be also crossed starting from the unscreened 
phase at particle-hole symmetry and rising $\ep_d$, namely, by doping. In Fig.~\ref{mu T} 
we show how the parameter $\mu$ behaves as function of $\ep_d$ from 0 
up to the critical point $\ep_d\simeq 0.26$ at $U=2$ and almost zero temperature.\\

\begin{figure}[hbt]
\centerline{\includegraphics[width=0.45\textwidth]{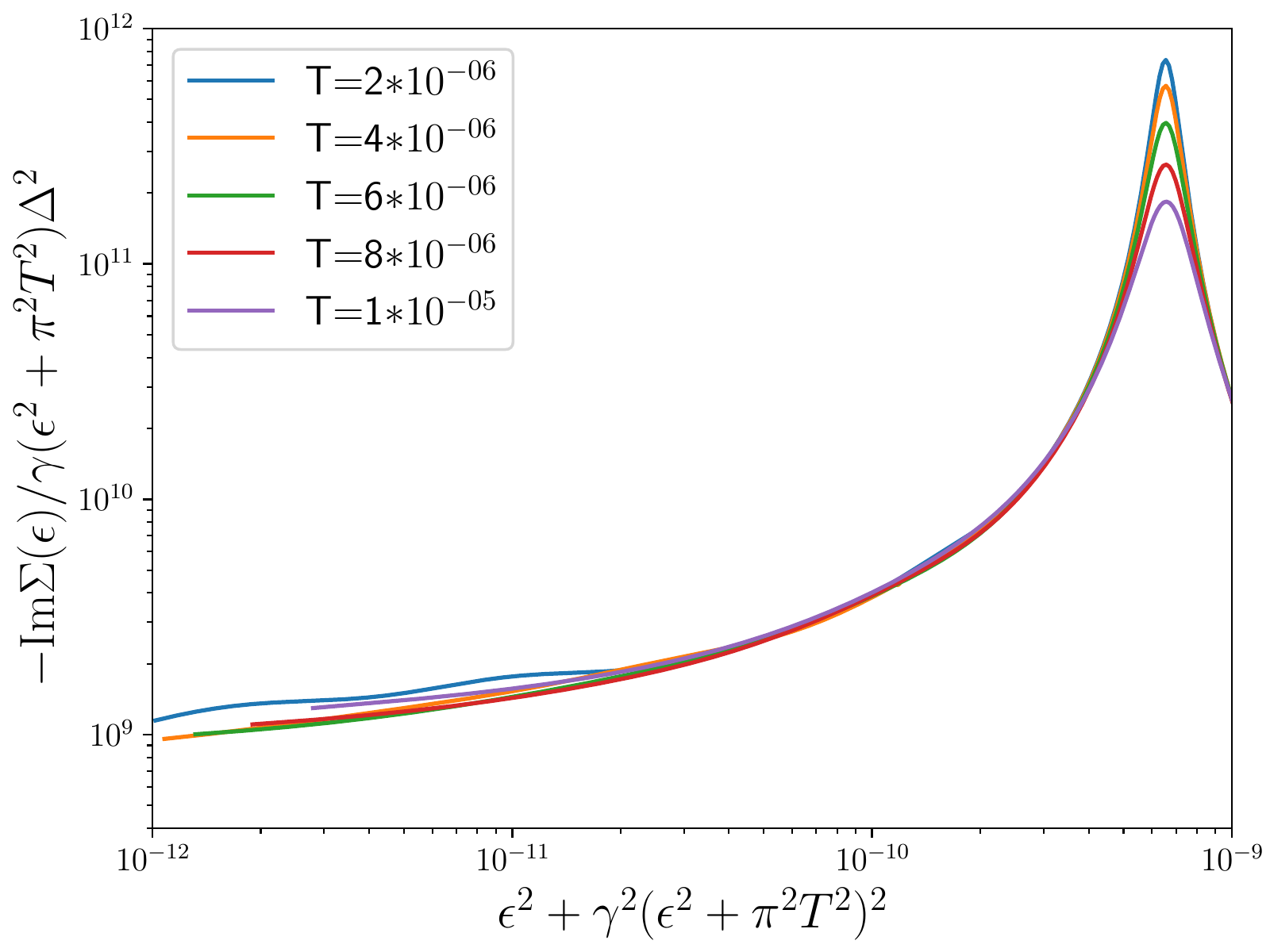}}
\caption{ Imaginary part of the self-energy in Eq.~\eqn{Sigma-no-ph} divided by $\Delta^2\,\gamma\,\big(\ep^2
+\pi^2\,T^2\big)$ as function of $\big(\ep^2+\pi^2\,T^2\big)$ for different $T$ at $\ep_d=0.1$ and $U=2$. 
Note the collapse of all curves at low Matsubara frequencies.}
\label{fit-ima}
\end{figure}

We remark that a key feature of the self-energy \eqn{fit} is the imaginary part in the denominator, i.e., $\gamma\big(\ep^2+\pi^2\,T^2\big)$, vanishing quadratically for $\ep,T\to 0$, see Fig.~\ref{fit-ima}.
This guarantees the existence of a well-defined `quasiparticle' excitation, 
namely whose decay rate at $T=0$ 
\beal
\gamma(\ep) &\equiv -Z(\ep)\,\Ima\,\Sigma_+(\ep)\propto \ep^2\,,
\eal 
with
\beal
Z(\ep) = \left(1-\fract{\partial\Rea\,\Sigma_+(\ep)}{\partial\ep}\right)\,,
\eal
vanishes at zero energy, even in the singular case at p-h symmetry~\cite{mio,mio-2}.
Such property distinguishes Eq.~\eqn{fit} from all model self-energies introduced to describe the pseudo-gap 
phase of underdoped cuprates, where either the imaginary part is missing, or assumed to be constant. 
We believe that Eq.~\eqn{fit}, though referring to a specific impurity model, is actually representative of 
generic pseudo-gap metal phases, and thus can be regarded as paradigmatic of such physical systems. 

\section{Luttinger's theorem in a non-magnetic Mott insulator}
\label{Rosh}
We now discuss the failure of Luttinger's theorem in the non-magnetic two-orbital Mott insulator analysed in Ref.~\cite{Rosh-2007}. The model is essentially the bulk generalisation of the 
two-orbital Anderson impurity model \eqn{HAIM}, with $U$ and $J$ much larger than 
the width $W$ of the conduction band, whose dispersion is assumed to be 
$\ep_{ij}(\bk)=\delta_{ij}\,\ep_\bk$, with $i,j=1,2$ the orbital indices. In that limit and for small enough deviation $\mu$ from particle-hole symmetry, the ground state of the model represents a non-magnetic Mott insulator 
with two electrons per site, each in a different orbital, locked into a spin singlet. 
The spin and orbital independent Green's function is 
\beal
G(i\ep,\bk) &= \fract{1}{\;i\ep -\ep_\bk+\mu-\Sigma(i\ep,\bk)\;}\;,\label{G-Rosh}
\eal
where, absorbing the Hartree-Fock self-energy in $\mu$, and to leading order in the hopping~\cite{Rosh-2007}, 
\beal
\Sigma(i\ep,\bk) &\simeq \fract{\Delta^2}{i\ep+\mu}\;,\label{Sigma-Rosh}
\eal
with $2\Delta= U+6J$ for the Hamiltonian \eqn{HAIM}. Rigorously speaking, 
the expression \eqn{Sigma-Rosh} is valid if also $|\mu| \gg W$, otherwise 
additional $\bk$-dependent terms appear in the denominator~\cite{Rosh-2007}. 
Therefore, we hereafter assume consistently that 
\beal
U\,,J\, ,|\mu| \gg W\,,\label{Rosh-assume}
\eal
which also implies that a Luttinger surface is absent. 
The Green's function \eqn{G-Rosh} describes an insulator lacking a Fermi surface if 
\beal
\Rea\,G(0,\bk)^{-1}=G(0,\bk)^{-1} &= -\ep_\bk+\mu -\fract{\Delta^2}{\mu}\;,
\eal
never vanishes within the Brillouin zone. That implies either 
$0<\mu<\mu_+$, in which case $G(0,\bk)^{-1}<0$, or $\mu_-<\mu<0$, 
in which case $G(0,\bk)^{-1}>0$, where~\cite{Rosh-2007}
\be
\mu_++\mu_- = \ep_\bk\,,\quad
\mu_+-\mu_- = \sqrt{\ep_\bk^2 + 4\Delta^2\;}\simeq 2\Delta\,.
\ee  
If we use conventional Luttinger's theorem, according to which the number of electrons per site 
is simply $n = 2 + 2\,\text{sign}\big( G(0,\bk)^{-1}\big)$, we obtain the wrong result that $n=4$ if $\mu_-<\mu<0$ 
and $n=0$ if $0<\mu<\mu_+$. However, in this case we can  
use Eq.~\eqn{PT-NO-AIM} to calculate the correct electron number. 
Indeed, since 
\beal
\lim_{\ep\to\infty}\Rea\,G(i\ep,\bk)^{-1} 
&= \mu-\ep_\bk \simeq \mu\,, 
\eal
the $\Rea\,G(i\ep,\bk)$ crosses zero an odd number of times 
from $\ep=0$ to $\ep=\infty$. According to Eq.~\eqn{PT-NO-AIM}, valid for a local self-energy close to half-filling, that implies $n_{1\bk\sigma}+n_{2\bk\sigma} =1$, which is indeed correct. \\
If the condition $|\mu|\gg W$ is not fulfilled, the no more negligible momentum dependent terms in the denominator of the self-energy Eq.~\eqn{Sigma-Rosh} yield a true Luttinger surface for a small interval of $\mu$ around zero~\cite{Rosh-2007}, 
in which case Eq.~\eqn{Oshikawa result} provides the correct electron number.

\subsection{Atomic limit of the $SU(N)$ Hubbard model}
\label{SU(N)}
At $J=0$ the previous model becomes the $N=4$  
$SU(N)$ Hubbard model at half-filling, which admits, for strong enough $U$,  
a Mott insulating state at any integer density $n=1,\dots,N-1$. In the atomic limit, $W =0$,  
also this model strongly violates Luttinger's theorem~\cite{Phillips-PRL2013}.  
However, the ground state in the atomic limit has an extensive degeneracy, 
$\binom{N}{n}$ per site, and thus divergent susceptibilities. In this situation, one does not expect Luttinger's theorem to apply~\cite{Igor-PRB2003,Potthoff-2006}.\\
Nonetheless,  to make a connection with the previous discussion, we note that the sum of the $N$ local Green's functions in the atomic limit at $T=0$~\cite{Phillips-PRL2013},
\beal
N\,G(i\ep) &= \fract{n}{\;i\ep+\ep_-\;}+\fract{N-n}{\;i\ep-\ep_+\;}\\
&= \fract{\partial}{\partial i\ep}\,\ln\big(i\ep+\ep_-\big)^n\,
\big(i\ep-\ep_+\big)^{N-n}\\
&\equiv -\fract{\partial \ln G_N(i\ep)}{\partial i\ep}\,,
\label{G-SU(N)}
\eal 
where $\ep_+=Un-\mu > 0$ and $\ep_-=\mu-U(n-1)>0$ are, respectively, the energies for adding and removing 
an electron from the atomic $n$-particle ground state. 
Therefore,
\bealn
N\,G(i\ep)\,\fract{\partial\Sigma(i\ep)}{\partial i\ep} &= 
N\,G(i\ep) + \fract{\partial\ln G(i\ep)^N}{\partial i\ep} \\
&= \fract{\partial}{\partial i\ep}\,\ln \fract{\;G(i\ep)^N\;}{G_N(i\ep)}\;,
\eal
is consistent with Eq.~\eqn{I_L-deriv} and quantised. In this case, it trivially follows that the role of the Luttinger 
integral is to freeze the occupation per orbital at $n/N$ rather than the value predicted by Luttinger's theorem, which 
is either 0 or $1$~\cite{Phillips-PRL2013} depending on $\mu$, in that similar to what we have found close to a half-filled Mott insulator. This suggests a natural extension of our results to multi-band models 
close to a Mott insulator at fractional filling $n/N$. \\
We end observing that $G_N(i\ep)$ is equivalent to the determinant of the $N\times N$ Green's function matrix 
corresponding to the same Hamiltonian but in presence of an infinitesimally 
small symmetry breaking field that lowers  
$n$ orbitals with respect to the other $N-n$ ones. In this case, Luttinger's 
theorem does hold, as Logan \textit{et al.} have explicitly demonstrated in the simpler $SU(2)$ case~\cite{Logan-JPC2015}.


\section{Discussion}
We have shown that the Luttinger integral, which provides the missing contribution to the electron count when 
Luttinger's theorem is violated, is a boundary zero-energy term and it is quantised in integer values when the 
self-energy is analytic at any non-zero imaginary frequency. 
Specifically, in a periodic single-band model of interacting electrons, 
Luttinger's theorem is violated when perturbation theory breaks down and a Luttinger surface appears in the Brillouin zone. Taking properly into account the quantised contribution from the Luttinger surface, we have found that the volume fraction of the Fermi pockets only measures the doping fraction away from half-filling 
rather than the full filling fraction. \\
In addition, a by-product of our derivation of Luttinger's theorem is the prediction that quasiparticles do exist even in 
half-filled non-symmetry-breaking Mott insulators, provided they possess a Luttinger surface, thus extending the results of  Ref.~\cite{mio-2} to the case of a hard gap. We emphasise that our formal construction in Sec.~\ref{Luttinger's theorem}
just relies on the assumption \eqn{LT: condition quasiparticles}, with no reference to a model Hamiltonian. 
However, the analogy with so-called $U(1)$ spin-liquid insulators~\cite{Moessner-PRL2001,Motrunich-PRL2002,Fisher&Balents-PRB2004,Balents-PRL2008,Senthil-PRX2016} is self-evident, and 
suggests that the quasiparticles are actually spinons, and the Luttinger surface their  'Fermi' surface.



\begin{acknowledgments}
We acknowledge helpful discussions with Alessandro Toschi.  
This work received funding from the European Research Council (ERC) under the European Union's Horizon 2020 research and innovation programme, Grant agreement No. 692670 ``FIRSTORM''. 
\end{acknowledgments}

\bibliographystyle{apsrev4-2}
\bibliography{mybiblio}

\end{document}